\documentclass[twocolumn,usenatbib]{mnras}

\usepackage{amsbsy}
\usepackage{amssymb}
\usepackage{amsmath}
\usepackage{bm}

\usepackage{graphicx}
\input{epsf}

\usepackage{color}

\def\spose#1{\hbox to 0pt{#1\hss}}

\def\lta{\mathrel{\spose{\lower 3pt\hbox{$\mathchar"218$}}
     \raise 2.0pt\hbox{$\mathchar"13C$}}}
\def\gta{\mathrel{\spose{\lower 3pt\hbox{$\mathchar"218$}}
     \raise 2.0pt\hbox{$\mathchar"13E$}}}
     
\newcommand{\be}{\begin{equation}}
\newcommand{\ee}{\end{equation}}
\def\bea{\begin{eqnarray}}
\def\eea{\end{eqnarray}}

\newcommand{\veps}{\varepsilon}
\def\mut{\widetilde{\mu}}

\def\X{\mathrm{x}}
\def\Y{\mathrm{y}}
\def\B{\mathrm{B}}
\def\ch{\mathrm{c}}
\def\f{\mathrm{f}}

\def\x{{\rm x}}
\def\y{{\rm y}}

\def\n{{\rm n}}
\def\p{{\rm p}}

\def\s{{\rm s}}

\def\wid{w_i^{\Y\X}}
\def\wju{w^j_{\Y\X}}
\def\wjd{w_j^{\Y\X}}

\begin{document}


\title{Dynamical tides in superfluid neutron stars}

\author[Passamonti, Andersson and Pnigouras]{%
    A. Passamonti$^1$, N. Andersson$^2$ and P. Pnigouras$^{3,4}$ \\ \\
    $^1$ Via Greve 10, 00146, Roma, Italy \\
    $^2$ School of Mathematics and STAG Research Centre, University of Southampton, Southampton SO17 1BJ, UK \\
    $^3$ Dipartimento di Fisica, ``Sapienza" Universit{\`a} di Roma \& Sezione INFN Roma1, Piazzale Aldo Moro 2, 00185 Roma, Italy \\
    $^4$ Department of Physics, Aristotle University of Thessaloniki, 54124 Thessaloniki, Greece
}

\date{\today}

\maketitle


\begin{abstract}
    We study the tidal response of a superfluid neutron star in a binary system, focussing on Newtonian models with superfluid neutrons present throughout the star's core and the inner crust. 
    Within the two-fluid formalism, we consider the main  aspects 
    that arise from the presence of different regions inside the star, with particular focus on the various interfaces. Having established the relevant theory, we determine the tidal excitation of the most relevant oscillation modes during binary inspiral.
    Our results suggest that superfluid physics has a negligible impact on the static tidal deformation. 
    The overwhelming contribution to the Love number is given by, as for normal matter stars, the ordinary fundamental mode (f-mode).  Strong entrainment, here described by a phenomenological expression which mimics the large effective neutron mass expected at the bottom of the crust, is shown to have significant impact on the superfluid modes, but  our results for the dynamical tide are nevertheless similar to the static limit: the fundamental modes are the ones most significantly excited by the tidal interaction, with the ordinary f-mode 
    dominating the superfluid one. We also discuss the strain built up in the star's crust during binary inspiral, showing that the superfluid f-mode may (depending on entrainment) reach the limit where the crust breaks, although it does so after the ordinary f-mode. Overall, our results suggest that the presence of superfluidity may be difficult to establish from binary neutron star gravitational-wave signals.
\end{abstract}



\section{Context: Superfluid neutron star dynamics}

Mature neutron stars exhibit states of matter commonly associated with low-temperature physics. 
The main difference from the laboratory setting is that, for neutron stars, the relevant `reference temperature' (the Fermi temperature of the nucleons) is of order $10^{12}$~K.  Shortly after a neutron star is born, the outer layers freeze to form an elastic crust of neutron-rich  nuclei. Meanwhile, beyond the neutron-drip density the neutrons form a superfluid and in the star's fluid core the protons condense into a superconductor. More exotic phases of matter, like superfluid hyperons and/or colour superconducting quarks may come into play above the nuclear saturation density. In essence, astrophysical observations allow us to explore the hottest superfluids/superconductors in the Universe \citep{page,shtern1,shtern2}.

The natural question to ask is to what extent we can use neutron-star observations to establish, beyond the level of theoretical expectation, the actual presence of macroscopic superfluidity. This is clearly a challenge given that neutron stars represent `hands-off laboratories'. However, it is well established that the macroscopic glitches seen in the spin evolution of many young neutron stars (with the Vela pulsar the poster example) provide smoking-gun evidence \citep{bryn}. The rapid spin-up and gradual relaxation associated with a glitch is explained in terms of angular momentum transfer from a loosely coupled superfluid component. The argument is fairly intuitive, involving the unpinning of an ensemble of quantised vortices by means of which the superfluid mimics bulk rotation, but we do not yet have truly quantitative models that can be matched against observed events (although note the recent constraints on the glitch rise time in the Vela pulsar \citep{palf,grab,ash}).  

Superfluidity also enters discussions of neutron-star seismology. As an example, it is expected that the vortex mutual friction provides the dominant damping mechanism for the gravitational-wave driven instability of the star's fundamental oscillation mode \citep{fmode1,fmode2}. Many different aspects on superfluid neutron star oscillations have been explored since the seminal work of Mendell in the early 1990s \citep{mend1,mend2}. To cut a long story short, one of the main features of the superfluid problem is the doubling of (many of) the star's oscillation modes \citep{epstein,Lee,AC01,comer}. The standard two-fluid description involves two coupled components with the presence of an additional dynamical degree of freedom enriching the oscillation spectrum. Another natural question then is if there are situations where the additional modes come into play, perhaps in a way that leads to observationally distinguishable features? This is the context for the present discussion. 

We consider the tide raised in a superfluid star by a binary companion (treated as a point particle, which should be a good enough approximation for our purposes, see \citet{PWbook} for a detailed discussion of the formulation of  tidal problem). The tidal interaction induces a linear response in the primary, with both a static and a dynamical component. Both features may be described in terms of a sum over the star's oscillations modes, a fact that is well established for Newtonian single fluid stars (see \citealt{ap20a} and \citealt{PAP21} for recent discussions) and which we extend to superfluid models here. As the mode spectrum depends on the superfluid parameters, it seems reasonable to ask what the impact on the tidal  response may be. Previous work in this direction has considered the tidal deformability for two-fluid stars in relativity \citep{Datta}, the impact of internal differential rotation on the so-called $I$-Love-$Q$ relations \citep{Yeung} and the shift of the dynamical g-mode resonances due to superfluidity \citep{wein}. The latter effort is particularly relevant for the present discussion as it concerns the dynamical tidal response. However, in that case the focus was on a somewhat indirect feature of a superfluid star. The standard two-fluid model does not allow for g-modes due to composition gradients \citep{Lee, AC01} (see \citet{RW18} for a recent discussion of superfluid g-modes), often considered in tidal discussions. The g-modes are reinstated in models that allow for muons (which should appear beyond a moderate density). This then leads to an electron-muon composition variation and an associated set of g-modes \citep{GK13, KG14, PAH16}. These new g-modes lead to tidal resonances appearing at higher frequencies than in the corresponding single-fluid model. However, indications are that this has little impact on the detectability of the contribution to dynamical tide---which would be a distinct signature of superfluidity---as the resonances appear at the late stage of binary inspiral where the system evolves too rapidly to significantly excite the modes \citep{wein}. In this paper, our main focus is on the role of the second degree of freedom that leads to additional oscillation modes, a direct feature of superfluidity. As this problem has not previously been explored in detail, we develop the relevant formalism (aspects of which are also discussed by \citealt{kirsty} and \citealt{wein}) making contact with, in particular, the discussion of superfluid oscillations from \citet{pass} and highlighting the impact of the entrainment effect on the two-fluid dynamics.


\section{Recap: The single-fluid  problem}

In order to provide context for the discussion, and establish some of the key results we will need, let us remind ourselves of the results for single-fluid systems.  If we want to quantify the fluid response to a tidal driving potential---in essence, quantify the dynamical tide---we need to solve the linearised fluid equations (here in Newtonian gravity). Focussing on non-rotating stars and following \citet{fs78}, we express the perturbations in term of the Lagrangian displacement vector (with components) $\xi^i$, governed by
\be
A_i^{\ j} \partial^2_t \xi_j + C_i^{\ j} \xi_j = -\rho \nabla_i \chi \ ,
\ee
where $\chi$ is the tidal potential associated with the binary companion. We simply have $A_i^{\ j}=\rho \delta^j_i$, with $\rho$ the density of the unperturbed star, while the expression for the $C_i^{\ j}$ operator is more involved (but we will not need it here).

Assuming a mode expansion 
\be
\xi^i = \sum_n a_n(t) \tilde \xi_n^i (r,\theta,\varphi) \ ,
\label{modexp}
\ee
where the time-independent mode eigenfunctions $\tilde \xi_n^i$ are solutions to the corresponding homogeneous problem (including the relevant boundary conditions), and that the modes behave as $e^{i\omega_n t}$, we have 
\be
-\omega_n^2  A^i_{\ j} \tilde \xi_n^j + C^i_{\ j} \tilde \xi_n^j = 0  \ .
\ee
As demonstrated by \citet{fs78},  two mode solutions (labelled $n$ and $n'$) are orthogonal in the sense that 
(here, and in the following, suppressing the spatial indices for clarity)
\be 
\langle \tilde \xi_{n'}, \rho \tilde \xi_n\rangle = \int \rho \tilde \xi_{n'}^* \tilde \xi_n dV = \mathcal A^2_n \delta _{nn'}  \ ,
\ee
where $\mathcal A_n^2$ is a normalisation constant. Making use of this we have
\begin{multline}
 \langle \tilde \xi_{n'},\sum_n \left( \ddot a_n + \omega_n^2 a_n \right) \rho \tilde \xi_n \rangle \\
 =    \sum_n \left( \ddot a_n + \omega_n^2 a_n \right) \langle \tilde \xi_{n'},\rho \tilde \xi_n \rangle \\
 = \left( \ddot a_{n'} + \omega_{n'}^2 a_{n'} \right) \mathcal A^2_{n'} = - \langle \tilde \xi_{n'},\rho \nabla \chi \rangle 
\end{multline}
or
\be
\ddot a_{n} + \omega_{n}^2 a_{n} = - \frac{1}{\mathcal A^2_{n}} \langle \tilde \xi_{n},\rho \nabla \chi \rangle \ .
\label{aneq0}
\ee
 Once we account for the time dependence of the tidal potential, we can solve this equation for the individual mode amplitudes $a_n$. The usual procedure involves working in the frequency domain, but if we assume that the binary orbit evolves adiabatically we can make a short cut. 
We have
\begin{multline}
 \chi(t,r,\theta,\varphi) =- GM'\sum_{l\ge2} \sum_{m=-l}^l  W_{lm} {r^l  Y_l^m    \over D^{l+1}} e^{-im\Omega t} \\
 =  \sum_{l\ge2} \sum_{m=-l}^l  \hat \chi_{lm}   Y_l^m    e^{-im\Omega t} 
 \ ,
\label{hatchi}\end{multline}
where $M'$ is the mass of the binary companion, $\Omega$ is the orbital frequency and $D$ is the orbital separation. Assuming a circular orbit in the equatorial plane, the $W_{lm}$ coefficients can be found in, for example, \citet{Lai} and we provide specific examples later when we work out the tidal mode excitation.  As we are interested in the driven solution to \eqref{aneq0}, we see that $a_n$ must inherit the time dependence of the tidal potential. With (for a given value of $m$)
\be
a_n = \hat a_n e^{-im\Omega t}
\ee
we have
\be
\hat a_{n} = - {1\over \mathcal A^2_{n} \left[\omega_n^2 - (m\Omega)^2\right] } \langle \tilde \xi_{n},\rho \nabla \hat \chi \rangle \ . 
\label{aneq1}
\ee
Finally,  we introduce the inner product between the tide and the mode eigenfunction 
\begin{multline}
Q_n = -\langle \tilde\xi_n , \rho \nabla \hat \chi \rangle = -\int \rho \tilde \xi_n^{i*} \nabla_i \hat \chi dV \\
= \int \hat \chi \nabla_i (\rho \tilde \xi_n^{i*}) dV = - \int \hat \chi \delta \tilde \rho_n^* dV \ ,
\end{multline}
where we used the continuity equation for the (Eulerian) density perturbation $\delta\rho$ in the last equality.
Expanding the perturbed density in spherical harmonics, we are left with
\begin{multline}
Q_{nlm} = GM' \sum_l  {W_{lm} \over D^{l+1} } \int r^{l+2} \delta \tilde \rho_{nlm}^* dr \ ,
\\
\equiv GM' \sum_l  {W_{lm} \over D^{l+1} }  I_{nlm}^*
\end{multline}
where the mass multipole moment $I_{nlm}$ is also referred to as the ``overlap integral". Since we know that---because of the spherical symmetry---each mode  contributes only to a single $(l,m)$-multipole for a non-rotating star, we may dispose of the sums over $l$ and $m$. We then have
\be
\hat a_{n} =  {Q_{nlm}\over \mathcal A^2_{n}  \left[\omega_n^2 - (m\Omega)^2\right] }  \ .
\label{aneq}
\ee

We also need (integrating the perturbed Poisson equation by parts)
\begin{multline}
4\pi G \int_0^R r^{l+2} \delta \tilde \rho_{nlm} dr \\ = 
R^{l+2} \left[{d \over dr} \delta \tilde \Phi_{nlm} \right]_{r=R} 
-l  R^{l+1}  \delta \tilde \Phi_{nlm} (R) \ ,
\label{intpart}
\end{multline}
where we know that the solution should satisfy (assuming that the density vanishes at the star's surface, $r=R$) 
\be
{d \over dr} \delta  \tilde \Phi_{nlm} + {l+1 \over r} \delta  \tilde \Phi_{nlm} =0 \quad \mbox{at} \quad r=R \ . 
\label{potcon}
\ee
Hence, \eqref{intpart} leads to
\be
4\pi G \int_0^R r^{l+2} \delta  \tilde \rho_{nlm} dr = - (2l+1) R^{l+1}  \delta \tilde \Phi_{nlm} (R) \ , 
\ee
or, connecting to the mass multipole moment (noting the different sign convention from \citealt{ap20a}, which has no impact on the final result)
\be
I_{nlm}  =  - { 2l+1 \over 4\pi G } R^{l+1}  \delta  \tilde \Phi_{nlm}^* (R) \ .
\ee

Expanding the induced perturbation in the gravitational potential in terms of the mode solutions, we have
\be
\delta \Phi = \sum_{l,m} \delta \Phi_{lm}Y_l^m =  \sum_{l,m} \sum_n a_n \delta \Phi_{nlm}Y_l^m    \ .
\ee
We then have, at the star's surface,
\begin{multline}
\delta \Phi_{lm} = - 4\pi G \sum_n {1\over (2l+1) R^{l+1}} \\ \times {1\over \mathcal A^2_{n} [\omega_n^2 - (m\Omega)^2] }  Q_{nlm} I_{nlm}  e^{-im\Omega t} 
\end{multline}
which, as each mode depends on  single values of $l$ and $m$, leads to
\begin{multline}
\delta \Phi_{lm} = - { 4\pi G^2 M' W_{lm} \over (2l+1) (DR)^{l+1}} \\\times \sum_n  {1\over \mathcal A^2_{n} [\omega_n^2 - (m\Omega)^2] }  | I_{nlm} |^2  e^{-im\Omega t} \ .
\label{dP}
\end{multline}

From before, we know that  (again, at the surface)
 \be 
 \chi_{lm} = - GM' W_{lm} {R^l \over D^{l+1}}   e^{-im\Omega t} \ .
 \label{cm}
 \ee
 Comparing \eqref{dP} to \eqref{cm} we see that 
 \be
 \delta \Phi_{lm} = 2 k_{lm} \chi_{lm}
\ee
leads to the (effective) Love number
\be 
k_{lm} =  {2\pi G \over (2l+1) R^{2l+1}} \sum_n  {1\over \mathcal A^2_{n} [\omega_n^2 - (m\Omega)^2] }  | I_{nlm} |^2 \ .
\ee
This expression for $k_{lm}$ is notably simpler than the, formally equivalent, result from \citet{ap20a}.

Finally, we introduce the dimensionless frequency $\tilde\omega_n$
\be
\omega_n^2 = \tilde \omega_n^2 \left( {GM_\star \over R^3} \right) 
\ee
and similar for $\tilde \Omega$, along with the normalisation 
\be
\mathcal A_n^2 = M_\star R^2
\ee
and
\be
\tilde I_n = {I_{nlm} \over M_\star R^l} \ ,
\ee
where $M_\star$ is the star's mass, to get the final result
\be 
k_{lm} =  {2\pi \over (2l+1)} \sum_n  {|\tilde I_n|^2 \over  \tilde \omega_n^2 - (m \tilde \Omega)^2 }  \ .
\label{modesum}
\ee
In the static limit, when $\tilde \Omega \to 0$, this result provides an expression for the Love number in terms of the set of modes of the star (notably degenerate in $m$ as one would expect). We can sanity check this by making use of the numerical results (for simple polytropic models) from \citet{ap20a}. An illustrative example is provided in Table~\ref{table1}, showing that the mode-sum is completely dominated by the contribution from the fundamental f-mode. This is as expected. What is perhaps remarkable is how small the contribution from the other modes is. For $\tilde \Omega \neq 0$ Eq.~\eqref{modesum} provides a simple expression for the dynamical tide, reflecting its origin in a set of driven stellar oscillation modes.

\begin{table}
\begin{tabular}{l c c c c}
\hline
mode & $\tilde \omega_n$ & $|\tilde I_n|$ & contribution to $k_2$ \\ 
\hline
$f$ 	& 1.2267 		& $5.5792\times 10^{-1}$ 		& 0.25994 \\ 
$p_1$ 	& 3.4615  		& $2.6888\times 10^{-2}$ 	    & $7.582\times 10^{-5}$ \\
$p_2$ 	& 5.4144 		& $2.6168\times 10^{-3}$ 		& $2.935\times 10^{-7}$ \\
$p_3$ 	& 7.2581  		& $3.0791\times 10^{-4}$ 		& $2.262\times 10^{-9}$ \\ 
$p_4$ 	& 9.0471  		& $4.4927\times 10^{-5}$ 		& $3.099\times 10^{-11}$ \\
\hline
\end{tabular} 

\caption{Sanity check of the mode-sum \eqref{modesum} (in the $\tilde \Omega \to 0$ limit) for the static Love number for barotropic $n=1$ polytropes (with an adiabatic index $\Gamma_1=2$) and $l=2$ (a model which only exhibits the f-mode and a set of higher frequency p-modes).  The sum ($\approx 0.26001$) should be compared to the expected result $k_2 = (15/\pi^2-1)/2\approx 0.259909$, see \citet{PWbook}. [Based on numerical mode results from \citet{ap20a}].}
\label{table1}
\end{table}


\section{The superfluid problem}
\label{sec:sfproblem}

In order to establish how superfluidity impacts on the tidal response we make use of the standard two-fluid model. That is, we separate the superfluid neutrons from the (co-moving) charged components (protons and electrons in the simplest case), and carefully consider the role of the two (coupled) fluid degrees of freedom in the problem. Formally, this involves repeating the steps from the single-fluid case. 
First, we need to establish the appropriate orthogonality condition for mode solutions and derive the mode-sum formula for the dynamical tide. Second, we need to take a closer look at the fine print details of the problem. This involves considering  the junctions conditions at the superfluid transition density, how the entrainment enters the problem and the implication of the presence of a  solid crust (including the conditions that must be satisfied at the crust-core interface). We have quite a lot of ground to cover, so let us get going.


\subsection{Representing the two degrees of freedom}

As in the single-fluid problem, we begin by deriving the 
equations governing Lagrangian perturbations of the system. In the first instance ignoring the entrainment coupling---in order to keep the discussion as transparent as possible---we have the Euler equations \citep{kirsty}
\begin{eqnarray}
  \left( \partial_t   + \mathcal L_{v_\X}\right) v^\X_i   +  \nabla_i 
       \left(\Phi + \tilde{\mu}_\X - { 1\over 2} v_\X^2\right)
 = 0 \ ,
\label{eul1}\end{eqnarray}
where $\x$  represents one of the two fluid degrees of freedom (and  $\y\neq \x$ provides the other one later) and $\mathcal L_{v_\X}$ denotes the Lie derivative with respect to $v_i^\X$. As we focus on non-rotating stars (some of the reasons for this are explained later when we discuss the entrainment), the background equilibrium must be such that (since $v_\x^i=0$)
\begin{equation}
    \nabla_i \left( \Phi + \tilde \mu_\x\right) = 0 \ ,
    \label{sfequil}
\end{equation}
where we have introduced $\tilde \mu_\x = \mu_\x/m_\B$, with $\mu_\x$ the chemical potential of the x component and $m_\B$ the baryon  mass.

Clearly, we must now work with two distinct
Lagrangian displacement vectors ${\xi}^i_\X$ \citep{kirsty}. In order to distinguish between the two possibilities we use Lagrangian
variations $\Delta_\X$ such that
\begin{equation}
\Delta_\X Q = \delta Q + \mathcal L_{\xi_\X} Q \ .
\end{equation}
The  generalisation of the single-fluid results for co- and contravariant vectors is straightforward. 
The two Lagrangian variations are introduced
in such a way that
\begin{equation}
\Delta_\X v_\X^i = \partial_t \xi_\X^i
\end{equation}
which then leads to
the perturbed continuity equation for the number density $n_\x$ taking the form
\begin{equation}
\Delta_\X  n_\X = - n_\X \nabla_i \xi_\X ^i \quad \Longrightarrow
\quad \delta n_\X = - \nabla_i (n_\X \xi_\X^i) \ .
\end{equation}

With these definitions, it is  relatively easy to derive the perturbed 
Euler equations. From (\ref{eul1}) it follows that 
\begin{equation}
\partial_t^2 \xi^\X_i + 
\nabla_i \left( \delta \tilde \mu_\x+ \delta \Phi \right)
  = 0 \ ,
\label{sfpeul2}
\end{equation}
or
\begin{multline}
\partial_t^2 \xi^\X_i 
- \nabla_i \left[ \left( {\partial \tilde{\mu}_\X \over \partial n_\X}
 \right)_{n_\Y} \nabla_j (n_\X\xi_\X^j) + 
\left( {\partial \tilde{\mu}_\X \over \partial {n_\Y}} \right)_{n_\X} 
\nabla_j (
n_\Y\xi_\Y^j)
\right]  \\
+ 
 \nabla_i \delta \Phi 
  = 0 \ .
  \label{sfpeul3}
\end{multline}

From this result it is evident that  the two fluids are coupled even in the absence of entrainment. We need to consider the fact that they live in the same gravitational potential---gravity cannot distinguish neutrons from protons/electrons---and the equation of state may also lead to `chemical coupling', arising when the chemical potential of one species depends on the presence of particles of the other kind  \citep{mend1,mend2,Lee,AC01,PR02,GK11,Gua14,LivRev}.

The perturbed 
gravitational potential explicitly depends on both displacement vectors.
Labelling the two components by n (not to be confused with the mode label $n$) and p---later taken to represent the superfluid neutrons and a conglomerate of all other components in the neutron star core, respectively---we have
\begin{multline}
\nabla^2 \delta \Phi = 4\pi m_\B G ( \delta n_\n + \delta n_\p)  \\ =
- 4\pi m_\B G \nabla_i (n_\n\xi_\n^i + n_\p \xi_\p^i) \ .
\end{multline}
Since this is a linear equation we can write the solution as
\begin{equation}
\delta \Phi = \delta \Phi_\n + \delta \Phi_\p = \sum_{\X=\n,\p} 
\delta \Phi_\X
\end{equation}
where we define
\begin{equation}
\nabla^2 \delta \Phi_\X = 4\pi m_\B G  \delta n_\X  = - 4\pi m_\B G
\nabla_i (n_\X\xi_\X^i) \ .
\end{equation}

There are clearly different ways to write the perturbation equations, but
we  know from \citet{mend1,mend2,Lee,AC01} and \citet{PR02} that it makes sense to represent one of the degrees of freedom by an equation for the total momentum. In essence, we want to compare \eqref{sfpeul3} to the perturbed single-fluid equation.
In order to effect this comparison, we 
introduce $\rho_\x = m_\B n_\x$ along with the weighted displacement \citep{AC01} 
\begin{equation}
\rho \xi^i = \rho_\n \xi_\n^i + \rho_\p \xi_\p^i \ .
\label{total}
\end{equation}
Adding the two perturbed Euler equations, we then get
\begin{equation}
\partial_t^2 \xi_i  + {1\over \rho} \left( n_\n \nabla_i \delta \mu_\n + n_\p \nabla_i \delta \mu_\p \right)   + \nabla_i \delta \Phi = 0  \ .
\end{equation}
Here we have
\begin{equation}
\delta p = n_\n \delta \mu_\n + n_\p \delta \mu_\p
\end{equation}
and we also introduce \citep{AC01}
\begin{equation}
\delta \beta = \delta \tilde \mu_\p - \delta \tilde \mu_\n
\end{equation}
such that 
\begin{equation}
\delta \tilde\mu_\n = {1 \over \rho} \delta p - {\rho_\p \over \rho} \delta \beta
\label{dmun}
\end{equation}
and
\begin{equation}
\delta \tilde \mu_\p =  {1 \over \rho} \delta p + {\rho_\n \over \rho} \delta \beta \ .
\end{equation}
This then leads to
\begin{multline}
 {1\over \rho} \left( n_\n \nabla_i \delta \mu_\n + n_\p \nabla_i \delta \mu_\p \right) =  {1\over \rho} \left( \rho_\n \nabla_i \delta \tilde \mu_\n + \rho_\p \nabla_i \delta \tilde \mu_\p \right) \\
=   \nabla_i  \left( {\delta p \over \rho} \right) -  \nabla_i \left(  {\rho_\p \over \rho} \right)  \delta \beta  \\
=  \nabla_i  \left( {\delta p \over \rho} \right) - \left[ {\partial \over \partial p}  \left(  {\rho_\p \over \rho} \right)_\beta \nabla_i p\right]  \delta \beta 
\label{2fdp}
 \end{multline}
 which reproduces the result from \citet{AC01} (and also agrees with \citealt{wein}). 
 
 Before we proceed, it makes sense to pause and note that in the barotropic single-fluid case, {which is relevant for the static tide \citep{ap19}}, we have
 \begin{equation}
 - \left( {1 \over \rho^2} \nabla_i p \right) \delta \rho = \nabla_i \left( {1\over \rho} \right) \delta p \ .
 \end{equation}
  {In effect, if we were to impose} $\delta \beta = 0$ in the single-fluid region, as would be natural in the static limit, it is obvious that \eqref{2fdp} extends continuously across the superfluid transition. By enforcing this condition we  {assume} that the single-fluid region is barotropic and hence cannot support g-modes.  {This is appropriate for a discussion of the static tide---and the usual Love number---but not for the dynamical tide that is our main interest here. In the latter case, we implement the same junction conditions as \citet{pass}, see the discussion later.}

 The fact that the new term in the Euler equation for the perturbed (total) momentum involves the deviation from chemical equilibrium associated with the perturbation provides an intuitive idea of the way  the two-fluid nature of the problem impacts on the tidal response. The effect will be large on modes that induce a significant $\delta \beta$ while the result should be close to the  single-fluid one when this quantity remains small. Our main aim is to make this expectation precise. 
 
Before we turn to this issue, and in order to complete the derivation, 
the dynamical equation for the second degree of freedom is easily obtained from  \eqref{sfpeul3}.  Introducing the difference (note that \citealt{wein} use a slightly different definition)
\begin{equation}
    \psi^i = \xi_\p^i - \xi_\n^i
    \label{diff}
\end{equation}
we immediately get
\begin{equation}
\partial_t^2 \psi_i + {1\over m_B} \nabla_i \delta \beta = 0 \ .
\end{equation}
Intuitively, the relative motion is driven by the deviation from chemical equilibrium \citep{mend1,AC01}.

In summary, the new variables $\xi^i$ and $\psi^i$ (later referred to as co- and counter-moving, respectively) are motivated by the fact that the two degrees of freedom would decouple if the second term in \eqref{2fdp} were to vanish. In general, this is not the case, but (as we will see later in Fig.~\ref{fig:eig-f}) the oscillation modes of the system are still fairly well represented by this intuition.


\subsection{The static tide} 

Before we consider the  problem of dynamical tides in a superfluid star, it is natural to comment on the (simpler) problem of the static tide, usually expressed in terms of the tidal deformability. Essentially, this means that we consider the perturbation equations in the static limit. In the two-fluid case, we then have
\begin{equation}
   {1\over \rho}  \nabla_i \delta p - {\delta p \over \rho^2} \nabla_i \rho  - \left[ {\partial \over \partial p}  \left(  {\rho_\p \over \rho} \right)_\beta \nabla_i p\right]  \delta \beta + \nabla_ i \delta \Phi = 0 
\end{equation}
along with
\begin{equation}
\nabla^2 \delta \Phi = 4\pi G \delta \rho
\end{equation}
and
\begin{equation}
     \nabla_i \delta \beta = 0 \ .
\label{beteq}
\end{equation}
The key point here is that \eqref{beteq} is consistent with the condition $\delta \beta =0$,  {mentioned earlier}, which would represent the two fluids being in chemical equilibrium. 
As discussed by \citet{ap19}, one would expect the unperturbed system to be in equilibrium in the static limit (effectively, in the static limit the tidal driving is much slower than the timescale required for the system to reach equilibrium). As a result, the fluid is adequately described as a barotrope $p=p(\rho$). In the normal fluid case, there will be no g-modes while in the superfluid case we have $\delta \beta =0$ and the second fluid degree of freedom has no effect on the tidal response. The argument is quite general. In order to distinguish the presence of a superfluid component we need to consider dynamical aspects of the tide. This will be brought out by our numerical results.


\subsection{Modes and orthogonality}

Now consider the general problem.  In analogy with the single fluid problem (and following \citealt{kirsty}), we can write the perturbed
Euler equations 
in the schematic form, after multiplying Eq.~(\ref{sfpeul3}) by $n_\X$, 
\begin{equation}
A_\X \partial_t^2 \xi_\X + C_\X \xi_\X + D_\X \xi_\Y = 0
\ .
\label{sfeom}\end{equation}
The first two terms are (fairly) obvious 
generalisations 
of the single fluid case, although they now relate to each of the two fluids. 
The last term is new, and describes the coupling between the 
components. Explicitly, it takes the form
\begin{equation}
D_\X \xi_\Y = -n_\X \nabla_i \left[\left( {\partial \tilde{\mu}_\X \over 
\partial
n_\Y} \right)_{n_\X} \nabla_j (n_\Y \xi_\Y^j) \right] + n_\X \nabla_i \delta 
\Phi_\Y \ .
\end{equation}

In order to develop a mode-sum approach for the tidal problem, we need to  introduce a suitable inner product. Given the single-fluid results, it is easy to show that we have the following 
symmetries  
\begin{eqnarray}
\left<\eta_\X,A_\X \xi_\X\right> &=& \left<\xi_\X,A_\X \eta_\X\right>^*   \ , \label{sfsymm1}\\
\left<\eta_\X,C_\X \xi_\X\right> &=& \left<\xi_\X,C_\X \eta_\X\right>^* \ , \label{sfsymm3}
\end{eqnarray}
where $\eta^i_\X$ can (at this point) be any solution to the perturbation equation for the x component. Integrating by parts, with $\hat n_i$ the outwards pointing normal to the surface and noting that (mixed second partial derivatives commute)
\begin{equation}
  \tilde \mu_\x = {1\over m_\mathrm B}\left( {\partial \mathcal E \over \partial n_\x} \right)_{n_\y} \Longrightarrow  {\partial \tilde \mu_\x \over \partial n_\y} =  {\partial \tilde \mu_\y \over \partial n_\x}
\end{equation}
where $\mathcal E$ is the energy representing the equation of state \citep{Prix}, we have
\begin{multline}
    - \int n_\x \eta_\x^{i*} \nabla_i \left[ \left( {\partial \tilde{\mu}_\X \over 
\partial
n_\Y} \right)_{n_\X} \nabla_j (n_\Y \xi_\Y^j) \right] dV \\
=  \int \left[n_\y\xi^i_\y \left( {\partial \tilde{\mu}_\Y \over 
\partial
n_\X} \right)_{n_\Y} \nabla_j (n_\x \eta^{j*}_\x)   - n_\x \eta_\x^{i*} \left( {\partial \tilde{\mu}_\X \over 
\partial
n_\Y} \right)_{n_\X} \nabla_j (n_\Y \xi_\Y^j) \right] \hat n_i dS \\
- \int  n_\Y \xi_\Y^i \nabla_i \left[ \left( {\partial \tilde{\mu}_\Y \over 
\partial
n_\X} \right)_{n_\Y}\nabla_j (n_\x \eta^{j*}_\x) \right] dV \ .
\end{multline}

For the gravitational potential we need 
\begin{multline}
\int n_\x \eta_\x^{i*} \nabla_i \delta_\xi \Phi_\y dV \\
= \int \left[ n_\x \eta_\x^{i*} \delta_\xi \Phi_\y \right] \hat n_i dS - \int  \delta_\xi \Phi_\y \nabla_i ( n_\x \eta_\x^{i*}) dV \\
   = \int
  \Bigg[  n_\x\eta_\x^{i*} \delta_\xi \Phi_\y - n_\y \xi^i_\y \delta_\eta \Phi^*_\x   \\
  + { 1 \over 4 \pi Gm_\B}  g^{ij} \left( \delta_\xi \Phi_\y   \nabla_j \delta_\eta \Phi_\x^* - \delta_\eta \Phi_\x^*   \nabla_j \delta_\xi \Phi_\y \right) \Bigg] \hat n_i dS \\
  + \int   n_\y \xi_\y^i \nabla_i  \delta_\eta \Phi_\x^* dV \ .
\end{multline}

Leaving out the surface terms for a moment, these results show that\footnote{In fact, the first term on the left is equal to the second term on the right, while the second term on the left balances the first term on the right, but all we require here is the balance of the sum of the two terms.}
\begin{equation}
  \left<\eta_\n,D_\n \xi_\p\right> + \left<\eta_\p,D_\p \xi_\n\right>  =  \left< \xi_\n, D_\n 
\eta_\p\right>^* + \left<\xi_\p, D_\p \eta_\n \right>^* \ . \label{sfsymm4}
\end{equation}
In order to work out what this means for the tidal problem, we first of all need to consider the orthogonality of the mode solutions. It is clear that a global mode must have the two components move with the same frequency. Consider two such solutions: first of all the set $(\xi_\n^i, \xi_\p^i)$ associated with frequency $\omega$ (momentarily omitting the mode label, for clarity) and secondly the set $(\bar\xi^i_\n, \bar\xi^i_\p)$ associated with $\bar \omega$.  Then we have 
\begin{multline}
0 = \langle \bar\xi_\n , -\omega^2 A_\n \xi_\n + C_\n \xi_\n + D_\n \xi_\p\rangle \\
= - \omega^2 \langle \xi_\n, A_\n \bar\xi_\n\rangle^* + \langle \xi_\n , C_\n\bar\xi_\n\rangle^* + \langle \bar\xi_\n , D_\n \xi_\p\rangle \\
=  - \omega^2 \langle \xi_\n, A_\n \bar\xi_\n\rangle^* +  \left(\bar \omega^*\right)^2\langle \xi_\n , A_\n\bar\xi_\n\rangle^* - \langle \xi_\n , D_\n \bar\xi_\p \rangle^* + \langle \bar\xi_\n , D_\n \xi_\p\rangle \\
= \left[  \left(\bar \omega^*\right)^2 -\omega^2\right] \langle \bar\xi_\n , A_\n\xi_\n\rangle - \langle \xi_\n , D_\n \bar\xi_\p \rangle^* + \langle \bar\xi_\n , D_\n \xi_\p\rangle
\end{multline}
which demonstrates that the individual component eigenfunctions are not orthogonal in the usual sense. However, in the same way we get for the other component
\begin{equation}
\left[  \left(\bar \omega^*\right)^2 -\omega^2\right] \langle \bar\xi_\p , A_\p\xi_\p\rangle - \langle \xi_\p , D_\p \bar\xi_\n \rangle^* + \langle \bar\xi_\p , D_\p \xi_\n\rangle = 0 \ .
\end{equation}
If we add the two relations and make use of \eqref{sfsymm4}, we have
\begin{multline}
\left[  \left(\bar \omega^*\right)^2 -\omega^2\right] \left(  \langle \bar\xi_\n , A_\n\xi_\n\rangle + \langle \bar\xi_\p , A_\p\xi_\p\rangle\right) \\
=   \langle \xi_\p , D_\p \bar\xi_\n \rangle^* - \langle \bar\xi_\p , D_\p \xi_\n\rangle +  \langle \xi_\n , D_\n \bar\xi_\p \rangle^* - \langle \bar\xi_\n , D_\n \xi_\p\rangle = 0 \ .
\end{multline}
It appears that, for modes with distinct (real) frequencies we must have
\begin{equation}
\langle \bar\xi_\n ,\rho_\n\xi_\n\rangle + \langle \bar\xi_\p , \rho_\p\xi_\p\rangle = 0 \ , \quad m\neq n \ .
\label{xiort}
\end{equation}
Now introduce the `sum and difference' displacements $\xi^i$ and $\psi^i$ from \eqref{total} and \eqref{diff}, respectively, to get
\begin{equation}
\rho \xi^i = \rho_\n \xi^i_\n + \rho_\p \xi^i_\p = \rho \xi_\n^i + \rho_\p \psi^i \Longrightarrow \xi^i_\n = \xi^i - x_\p \psi^i \ ,
\end{equation}
where $x_\p=\rho_\p/\rho$ is the proton fraction, and
\begin{equation}
\xi^i_\p = \xi^i + (1-x_\p) \psi^i \ .
\end{equation}
Use these results in \eqref{xiort} to get (where the meaning of $\bar \xi$ and $\bar \psi$ should be obvious)
\begin{multline}
\langle \bar \xi  - x_\p \bar \psi ,\rho_\n (\xi - x_\p \psi )\rangle + \langle \bar \xi+(1-x_\p) \bar \psi , \rho_\p (\xi+(1-x_\p) \psi )\rangle \\
= \langle \bar \xi, \rho \xi\rangle    \underbrace{- \langle x_\p \bar\psi, \rho_\n \xi \rangle + \langle\rho_\n \bar\psi, x_\p \xi\rangle}_{=0} \\
+  \underbrace{\langle x_\p \bar \psi,  \rho_\p (1-x_\p) \psi\rangle +  \langle (1-x_\p) \bar\psi, \rho_\p (1-x_\p) \psi \rangle}_{=\langle \bar \psi, \rho_\p (1-x_\p) \psi \rangle } \\
= \langle \bar \xi, \rho \xi\rangle + \langle \bar \psi, {\rho_\n \rho_\p \over \rho} \psi \rangle
 = 0 
 \label{sfnorm}
\end{multline}
This establishes the sense in which the superfluid mode solutions are orthogonal. It accords with the conclusions from \citet{kirsty} and \citet{wein}.


\subsection{The tidal response}

Let us now work out how the additional degree of freedom affects the  mode-sum argument for the tidal response. This turns out to be relatively straightforward. 
Adapting the logic from the single fluid problem, we need 
\begin{equation}
\left( \begin{array}{cc} \xi^i \\ \psi^i \end{array}\right) = \sum_n a_n (t) \left( \begin{array}{cc} \tilde \xi_n^i \\ \tilde \psi_n^i \end{array}\right)
\end{equation}
where $\tilde \xi^i_n$ is a time-independent mode solution for the total displacement from \eqref{total}, while $\tilde \psi^i_n$ is the corresponding difference defined in \eqref{diff}. 

It is easy to see that the tidal potential---introduced by replacing $\delta \Phi \to \delta \Phi + \chi$ in the Euler equations---explicitly enters only the equation for the co-moving degree of freedom associated with $\xi^i$.
Assuming, as before, that the mode frequency is $\omega_n$ we get
\begin{equation}
\sum_n (\ddot a_n + \omega_n^2 a_n)  \left( \begin{array}{cc}   \tilde \xi_n^i \\ \tilde \psi_n^i \end{array}\right) = \left( \begin{array}{cc} - \nabla^i \chi \\ 0 \end{array}\right) \ .
\end{equation}
Next, using the inner product along with  the normalisation implied by \eqref{sfnorm},
\begin{equation}
\langle \tilde \xi_{m}, \rho \tilde \xi_n \rangle + \langle \tilde \psi_{m}, {\rho_\n \rho_\p \over \rho} \tilde \psi_n \rangle = \mathcal A_n^2  \delta_{mn} \ ,
\end{equation}
it is easy to show that
\begin{equation}
\ddot a_n + \omega_n^2 a_n   = - {1\over \mathcal A_n^2} \langle \tilde \xi_n, \rho \nabla_i \chi \rangle \ .
\end{equation}
Formally, the result takes the same form as in the single-fluid case (cf. Eq.~\eqref{aneq0}). The main difference 
is associated with the implicit presence of the difference degree of freedom in the normalisation integral. 

The final steps are easy. From the definition of the $\tilde \xi^i_n$ it follows that 
\be 
\delta \tilde \rho_n = - \nabla_i \left( \rho \tilde \xi^i_n \right) \ ,
\ee
just as in the single-fluid case. As a result, the derivation of the mode-sum for the tide proceeds as before (as it involves only the gravitational potential and the total density perturbation). The expression for the dynamical tidal response \eqref{modesum} remains unchanged (as long as we work with a neutron star model that has a single fluid surface, see the discussion in \citealt{PAP21}). The numerical results will, of course, be different. Before we explore this, we need to consider a few points of detail. 


\subsection{Fine print: The superfluid transition}

In order to complete a superfluid model, we need to understand the conditions to be imposed at the superfluid transition density. Superfluid neutrons will be present from the neutron drip density and the pairing gap may also be such that the superfluid region ends at some higher density (also depending on the star's interior temperature). The question then is how--- {in a dynamical setting}---we deal with the transition from superfluidity (and two  degrees of freedom) to normal matter (and a single fluid component).

 {As in \cite{pass}, we impose at 
neutron drip a `free-surface' condition on the free superfluid chemical potential, $\Delta \mu_\f=0$, along with continuity of the radial component
of the total momentum $\rho \xi^r = \rho_\mathrm{c} \xi^r_\mathrm{c} + \rho_\f \xi^r_\f$.  These conditions
and the vanishing of the free superfluid density at  neutron drip guarantee the continuity of the pressure perturbation.}


In a realistic  model there will be two transitions: 1) at the base of the crust, where it is natural to change `chemical gauge' (see later discussion) and 2) at neutron drip, where the superfluid disappears. We need to work out what happens to the inner product (especially the surface terms arising from the integration by parts) at each of these phase transitions.

In view of this, let us  take a closer look at the surface terms from the integrals in the two-fluid problem.
Adding all terms associated with the chemical potentials, as required for the total momentum, we have (again with $\hat n_i$ the outwards pointing normal of the transition surface and also using $\delta_\xi$ to indicate the variation associated with $\xi^i$, and similarly for $\delta_\eta$ and $\eta^i$) 
\begin{multline}
\left[n_\n\xi^i_\n \left( {\partial \tilde{\mu}_\n \over 
\partial
n_\n} \right)_{n_\p} \nabla_j (n_\n \eta^{j*}_\n)   - n_\n \eta_\n^{i*} \left( {\partial \tilde{\mu}_\n \over 
\partial
n_\n} \right)_{n_\p} \nabla_j (n_\n \xi_\n^j) \right] \hat n_i  \\
+ \left[n_\p\xi^i_\p \left( {\partial \tilde{\mu}_\p \over 
\partial
n_\n} \right)_{n_\p} \nabla_j (n_\n \eta^{j*}_\n)   - n_\n \eta_\n^{i*} \left( {\partial \tilde{\mu}_\n \over 
\partial
n_\p} \right)_{n_\n} \nabla_j (n_\p \xi_\p^j) \right] \hat n_i \\
+ \left[n_\n\xi^i_\n \left( {\partial \tilde{\mu}_\n \over 
\partial
n_\p} \right)_{n_\n} \nabla_j (n_\p \eta^{j*}_\p)   - n_\p \eta_\p^{i*} \left( {\partial \tilde{\mu}_\p \over 
\partial
n_\n} \right)_{n_\p} \nabla_j (n_\n \xi_\n^j) \right] \hat n_i \\
+
 \left[n_\p\xi^i_\p\left( {\partial \tilde{\mu}_\p \over 
\partial
n_\p} \right)_{n_\n} \nabla_j (n_\p \eta^{j*}_\p)   - n_\p \eta_\p^{i*} \left( {\partial \tilde{\mu}_\p \over 
\partial
n_\p} \right)_{n_\n} \nabla_j (n_\p \xi_\p^j) \right] \hat n_i \\
= - n_\n \delta_\eta \tilde \mu_\n^* ( \xi^i_\n \hat n_i) - n_\p \delta_\eta \tilde \mu_\p^* ( \xi^i_\p \hat n_i ) 
\\
+ n_\n \delta_\xi \tilde \mu_n ( \eta^{i*}_\n \hat n_i) + n_\p \delta_\xi \tilde \mu_\p ( \eta^{i*}_\p \hat n_i ) \ .
\end{multline}
In terms of  the  weighted displacements \eqref{total} and \eqref{diff},  we get from the first two terms
\begin{multline}
  \rho_\n \delta_\eta  \mu_\n^* ( \xi^i_\n \hat n_i) + \rho_\p \delta_\eta  \mu_\p^* ( \xi^i_\p \hat n_i )  
\\
  = \delta_\eta p^*  (\xi^i \hat n_i) + (1-x_\p) \rho_\p \delta_\eta \beta^* (\psi^i \hat n_i) \ .
\end{multline}
Comparing to the surface term from the stratified single-fluid problem---which is relevant in the outer crust (as long as we ignore the elastic aspects) we see that the first term represents the (anticipated) continuity of the perturbed pressure.  {Moreover, as $x_\mathrm{c}\to 1$ and the other quantities take finite values,
the second term will vanish at the superfluid transition.} Finally, when it comes to the conditions on the gravitational potential it is easy to see that we end up with the same result---once we use the weighted displacement---as  in the single-fluid case (see \citealt{PAP21}). 


\subsection{Second fine print: Entrainment}

Adding important realism to the model, let us now account for  entrainment \citep{Prix04,YL03,KG11,LivRev}. This effect introduces an additional (non-dissipative) coupling mechanism to the two-fluid model, an essential aspect if we want to describe the dynamics of the neutron star crust,  where Bragg scattering off of the nuclear lattice makes the `free' neutrons less mobile than one might otherwise have expected \citep{Carter05,Ch05,Ch06}. This, in turn, impacts on the detailed oscillation modes and hence the tidal mode sum.

Formally, the entrainment 
encodes how the internal energy of the system depends on the relative velocity of the two fluids. Introducing the entrainment parameter $\alpha$ (encoding how the system's energy functional $\mathcal E$ depends on the relative velocity between the two components \citep{mend1,Prix,Prix04,Carter05,LivRev,NA21}) and 
$\varepsilon_\X= 2 \alpha/\rho_\X$ (following \citealt{Prix04}), the Euler equations can be written 
\begin{equation}
  \left( \partial_t + v_\X^j \nabla_j \right) \left(v^\X_i + \varepsilon_\X \wid \right) 
  + \nabla_i \left(\Phi + \mut_\X \right) + \varepsilon_\X
  \wjd \nabla_i v_\X^j = 0 \ ,
\end{equation}
where  
\begin{equation}
 w_{\y\x}^i = v_\y^i - v_\x^i
\end{equation}
represents the relative velocity between the  components. Phenomenologically, entrainment is crucial for the understanding of observed pulsar glitches \citep{crust1,crust2}---but for the present purposes the introduction of bulk rotation in the unperturbed configuration makes the analysis much less tractable. Fortunately, for the tidal problem it is reasonable to assume that the system is sufficiently old that the  stars have spun down to the extent that we may ignore rotational effects.  As any relative rotation should be a small fraction of the bulk rotation rate\footnote{The evidence for this from glitches, which represent a relative change in the spin rate of up to order $10^{-6}$ (likely reflecting the degree of differential rotation sustained by the superfluid in the system), is clear.},  it should be very safe to ignore background terms involving $w_{\y\x}^i$. This leaves us with a much simpler model. We get some idea of what may happen if we push the relative rotation to the extreme from the analysis of \citet{Yeung}, but we will not explore that regime here.

First of all, the continuity equations and the Poisson equation 
are not affected by entrainment. 
Second, the perturbed chemical potential terms also remain as before. In principle, $ \delta \tilde{\mu}_\X$ depends on the entrainment
since the chemical potential $\mu_\X$ follows from the
partial derivative of the energy functional, $\mathcal E$,  with respect to the  number density, $n_\X$. Since $\mathcal E$ depends on the 
entrainment the Eulerian variation of the chemical potential inherits this dependence. In general, we have
\begin{multline}
\delta \tilde{\mu}_\X = - \left( {\partial \tilde{\mu}_\X \over \partial n_\X}\right)_{n_\Y, w^2}\nabla_j \left(n_\X \xi_\X^j \right)
\\
- \left( {\partial \tilde{\mu}_\X \over \partial n_\Y}\right)_{n_\X, w^2}\nabla_j \left(n_\X \xi_\X^j \right)
+ \left( {\partial \tilde{\mu}_\X \over \partial w^2}\right)_{n_\X,n_\Y} \delta w^2 \ ,
\label{dmurot}
\end{multline}
where 
\begin{equation}
\delta w^2 = 2 \wjd \delta \wju \ .
\end{equation}
However, if the relative flow vanishes in the background then the last term in \eqref{dmurot} vanishes, as well, and we are left with the terms we have already accounted for. 

In essence, when we account for entrainment (for a co-moving background configuration), the Euler equations change to
\begin{equation}
(1-\varepsilon_\x) \partial^2_t \xi_\x + \varepsilon_\x \partial^2_t \xi_\y + C_\x \xi_\x + D_\x \xi_\y = 0 
\end{equation}
where only the first two terms are different from the non-entrained case.
Noting that \citep{Prix04}
\begin{equation}
    2 \alpha = \rho_\n \varepsilon_\n = \rho_\p  \varepsilon_\p \ ,
\end{equation}
it is easy to see that the equation for the weighted displacement remains as before---entrainment does not impact on the total momentum. For the mode orthogonality, we
need (a straightforward extension of the earlier argument)
\begin{multline}
\langle  \bar\xi_\n ,  \rho_\n (1-\varepsilon_\n) \xi_\n + \rho_\n \varepsilon_\n \xi_\p \rangle + \langle \bar\xi_\p , \rho_\p (1-\varepsilon_\p) \xi_\p - \rho_\p\varepsilon_\p \xi_\n \rangle \\
= \langle  \bar\xi_\n , (\rho_\n-2\alpha) \xi_\n +2 \alpha \xi_\p \rangle + \langle \bar\xi_\p , (\rho_\p -2\alpha) \xi_\p + 2\alpha \xi_\n \rangle = 0 \ , \; m\neq n \ .
\end{multline}
As before, we express this in terms of  $\xi^i_\n = \xi^i - x_\p \psi^i$ and 
$\xi^i_\p = \xi^i + (1-x_\p) \psi^i$,  leading to
\begin{multline}
    \langle \bar \xi - x_\p \bar \psi, (\rho_\n -2\alpha) (\xi - x_\p \psi) + 2\alpha (\xi+x_\n \psi)\rangle \\
    +\langle \bar \xi + x_\n \bar \psi, (\rho_\p - 2\alpha) (\xi+x_\n \psi)+2\alpha (\xi-x_\p \psi) \rangle  \\
= \langle \bar \xi , \rho \xi \rangle +
\langle \bar \psi , \left[ \rho_\p (1-x_\p) - 2\alpha \right] \psi \rangle \ .
\end{multline}
We may also use 
\begin{multline}
  \rho_\p (1-x_\p) - 2\alpha = {1\over \rho} \left[ \rho_\p \rho_\n- 2\alpha (\rho_\n + \rho_\p) \right] \\
  = {\rho_\n \rho_\p \over \rho}   \left( 1 - \varepsilon_\p - \varepsilon_\n \right) \ .
\end{multline}

Finally, taking the difference between the two Euler equations, we now have
\begin{multline}
(1-\varepsilon_\p) \partial_t^2 \xi^i_\p + \varepsilon_\p \partial_t^2 \xi^i_\n - (1-\varepsilon_\n) \partial_t^2 \xi^i_\n - \varepsilon_\n \partial_t^2 \xi^i_\p + \nabla_i ( \delta \tilde \mu_\p - \delta \tilde \mu_\n) \\
=  \left( 1 - \varepsilon_\p - \varepsilon_\n \right) \partial_t^2 \psi^i + \nabla_i \delta \beta = 0 \ .
\end{multline}

In summary, in order to account for the impact of entrainment on the tidal response mode sum, we need to change the normalisation condition to
\begin{equation}
\langle \tilde \xi_{m}, \rho \tilde \xi_n \rangle + \langle \tilde \psi_{m}, {\rho_\n \rho_\p \left( 1 - \varepsilon_\p - \varepsilon_\n \right)\over \rho} \tilde \psi_n \rangle = \mathcal A_n^2  \delta_{mn} \ . \label{entnorm}
\end{equation}
Everything else stays the same (although the mode frequencies and eigenfunctions are obviously affected by the entrainment).


\subsection{Third fine print: The crust-core interface}

As discussed by, for example, \citet{pass}, the superfluid in the neutron star crust requires a slightly different treatment. We describe the crust in terms of two components, labelled c for all confined nucleons (both neutrons and protons) associated with nuclei and the elastic crust and f for the free (superfluid) neutrons (replacing the charged conglomerate $\p$ and the neutrons $\n$ from the core, respectively). The distinction is important if we aim to account for the crust elasticity, which involves the nuclei associated with the elastic lattice (and hence both neutrons and protons). With our convention, we then have
\begin{equation}
 \left( 1 - \varepsilon_\ch \right)  \frac{ \partial^2 \xi _{i}^{\ch} }{\partial t ^2 }  + \varepsilon_\ch \frac{ \partial^2 \xi _{i}^{\f}
}{\partial t ^2 }+ \nabla_{i} \left( \delta \tilde \mu_{\ch} + \delta \Phi  \right)                                                =  \frac{1}{\rho_\ch} \nabla^{j} \sigma_{ij}   
\end{equation}
and 
\begin{equation}
 \left( 1 - \varepsilon_\f \right) \frac{ \partial^2 \xi _{i}^{\f}
}{\partial t ^2 } + \varepsilon_\f  \frac{ \partial^2 \xi _{i}^{\ch} }{\partial t ^2 } + \nabla_{i} \left( \delta \tilde \mu_{\f}  + \delta \Phi \right) = 0 \ ,
\end{equation}
where the elastic stress tensor is defined as
\begin{equation}
  \sigma_{ij} = \check \mu \left( \nabla_{i} \xi_{j}^{\ch} + \nabla_{j}
    \xi_{i}^{\ch} \right) - \frac{2}{3} \check \mu \left(
    \nabla^{k}\xi_{k}^{\ch} \right) \delta_{ij} \, , \label{eq:sigij}
\end{equation}
and the shear modulus is denoted by $\check \mu$. The equations for the neutron star core are the same as before. Effectively, we replace  the indices c
and f with p and n, respectively, and set $\check\mu$ to zero. 

The key point is that the model assumes different `chemical gauges' in the two regions (\citealp{Carter05,Ch05,Ch06}; see also \citealp{NA21} for a recent discussion). The need for this follows from the physics. It is easy to identify all neutrons in the core with the superfluid degree of freedom, but in the crust some of the neutrons are contained in the nuclei that form the elastic component. Hence, we need to make a distinction between the two cases. The alternative would be to continue to count all the neutrons also in the crust but this would then impact on the elastic stress tensor, which would depend on both displacement vectors. Our chosen prescription (which is the same as in \citealt{lagrange}) is natural as it simplifies the elastic contribution.
Within the two-fluid description, the difference can be expressed in terms of the
`chemical gauge' parameter $a_c$, defined through
\begin{equation}
n_{\ch} = a_{\ch} n_{\p} \quad \Longrightarrow \quad
n_{\f} = n_{\n} + \left( 1 - a_\ch \right) n_{\p} 
\end{equation}
where $n_{\f}$ and $n_{\ch}$ are, respectively, the number density of
the free neutrons and the confined nucleons (both neutrons and protons), while $n_{\n}$ and
$n_{\p}$ are the total number densities of neutrons and protons.
One can show that the neutron conjugate momentum and the
chemical potential are independent of the chemical
gauge choice as long
as $a_{\ch}$ is either held fixed or depends only on the nuclear charge
number $Z$.  This gauge independence is important for the
derivation of the crust/core junction conditions. 

We need
\begin{multline}
\mu_\f = \left( {\partial \mathcal E \over \partial n_\f }\right)_{n_\ch} \\
= 
\left({ \partial \mathcal E \over \partial n_\n}\right)_{n_\p} \left( { \partial n_\n \over \partial n_\f } \right)_{n_\ch} + \left( {\partial \mathcal E \over \partial n_\p}\right)_{n_\n} \left( { \partial n_\p \over \partial n_\f } \right)_{n_\ch} \\
= \mu_\n \left( { \partial n_\n \over \partial n_\f } \right)_{n_\ch} + \mu_\p \left( { \partial n_\p \over \partial n_\f } \right)_{n_\ch} 
\end{multline}
but we have
\begin{multline}
d\mathcal E = \left( {\partial \mathcal E \over \partial n_\f }\right)_{n_\ch} dn_\f + \left( {\partial \mathcal E \over \partial n_\ch }\right)_{n_\f} dn_\ch \\
= \left( {\partial \mathcal E \over \partial n_\f }\right)_{n_\ch} dn_\n + \left[ (1-a_\ch) \left( {\partial \mathcal E \over \partial n_\f }\right)_{n_\ch} + a_\ch \left( {\partial \mathcal E \over \partial n_\ch }\right)_{n_\f}  \right] dn_\p \\
+ n_\p \left[ \left( {\partial \mathcal E \over \partial n_\ch }\right)_{n_\f} - \left( {\partial \mathcal E \over \partial n_\f }\right)_{n_\ch}\right] da_\ch \ .
\end{multline}
It is easy to see that, if $da_\ch=0$ we have \citep{Carter05}
\begin{equation}
\mu_\f = \left( {\partial \mathcal E \over \partial n_\f }\right)_{n_\ch} = \left( {\partial \mathcal E \over \partial n_\n }\right)_{n_\p} = \mu_\n \ .
\end{equation}
In terms of the number density of nuclei $N$ and the charge number $Z$, we have
\begin{equation}
    N = {n_\p \over Z}
\end{equation}
as all protons are associated with the nuclei. Similarly, the confined baryon number is
\begin{equation}
    n_\ch = A N \ ,
\end{equation}
where $A$ is the atomic number of the nuclei. We see that this leads to 
\begin{equation}
a_\ch = {A\over Z} \ .
\end{equation}
Now assume that the equation of state provides $A=A(Z)$ in such a way that
\begin{multline}
d a_\ch =  a_\ch'  dZ = a_\ch' \left( {1\over N} dn_\p - {n_\p \over N^2} dN \right)\\
= Z a_\ch' \left(  {dn_\p \over n_\p}  - {dN \over N} \right) \ ,
\end{multline}
where the prime indicates a derivative with respect to $Z$.
It seems natural to assume that this will vanish as long as all protons are locked inside nuclei. The inferred invariance of the neutron chemical potential then leads to
\begin{equation}
    \delta \mu_\f = \delta \mu_\n
\end{equation}
at the crust-core interface.

As we are counting all protons on the two sides of the interface, it stands to reason that we should also expect continuity of the corresponding radial displacement
\begin{equation}
    \xi^r_\ch = \xi^r_\p
\end{equation}
and we also know (from the  {continuity of the baryon current}) that the weighted displacement must be continuous. That is, we have
\begin{multline}
    \xi^r = (1-x_\p) \xi^r_\n + x_\p \xi^r_\p = (1-x_\ch) \xi^r _\f + x_\ch \xi^r_\ch \\
    \Longrightarrow (1-x_\ch) \xi^r_f = (1-x_\p) \xi^r_\n + (x_\p - x_\ch)  \xi^r_\p 
\end{multline}
which is the condition  {also} implemented by \citet{pass}.

Finally, we need---in order to avoid a contribution from  surface terms---the continuity of
\begin{equation}
    (1-x_\p) \rho_\p \delta_\eta \beta^* (\psi^i \hat n_i ) \ .
\end{equation}
How is this going to work out? First, for $\psi$ we need
\begin{multline}
\xi^r_\ch-\xi^r_\f = \xi^r_\p - {1-x_\p \over 1-x_\ch} \xi^r_\n - {x_\p-x_\ch \over 1-x_\ch} \xi^r_\p \\ 
= {1\over 1-x_\ch} \left[ 1-x_\ch - x_\p + x_\ch\right] \xi^r_\p- {1-x_\p \over 1-x_\ch} \xi^r_\n \\
=  {1-x_\p \over 1-x_\ch}  \left( \xi^r_\p - \xi^r_\n \right)
\end{multline}
and
\begin{multline}
 \mu_\p-\mu_\n = \left[ (1-a_\ch) \left( {\partial \mathcal E \over \partial n_\f }\right)_{n_\ch} + a_\ch \left( {\partial \mathcal E \over \partial n_\ch }\right)_{n_\f}  \right] - \mu_\f \\
 = a_\ch \left( \mu_\ch - \mu_\f \right) = {x_\ch \over x_\p} \left( \mu_\ch - \mu_\f \right) \ .
\end{multline}
We have already assumed that $\delta a_\ch =0$, so this leads to 
\begin{equation}
\delta  \mu_\p- \delta \mu_\n = {x_\ch \over x_\p} \left( \delta \mu_\ch - \delta \mu_\f \right) \ ,
\end{equation}
which allows us to establish the continuity at the crust-core interface.  {In order to complete the model, we also} need the traction condition, as discussed by \citet{pass}. 


\section{The neutron star model} \label{sec:NS}

In order to quantify the impact of superfluidity on the tidal response (expressed in terms of the mode-sum \eqref{modesum}) we 
revisit the model considered by \citet{pass}. That is, we have  a Newtonian star with distinct superfluid constituents in the crust and the core, composition gradients and 
entrainment. The star has four different layers: a core, an inner and outer crust and an ocean. 
Superfluid neutrons are present in the core and in the inner crust, where they drip out of the atomic nuclei. 
These two regions are  described by the two-fluid model. In the star's core, the first dynamical degree of freedom is associated with 
the superfluid neutrons and the second describes the motion of a conglomerate of charged particles (protons, electrons, \dots). 
Dynamically, the charged components are assumed to move like a single fluid (a simplifying assumption that allows us to leave out electromagnetism, as there is no charge current).  
In the outer crust, up to the neutron drip transition, neutrons are tied up in the nuclei so this region can be described by a single component. 
The single component (although fluid rather than elastic) model is also used to describe the ocean. 
To identify the various constituents, in the fluid core we use  n for neutrons and p for the charge-neutral mixture. In the inner crust we change 
labels because only a fraction of the neutrons are able to move relative to the crust nuclei. We use    
f to represent the free neutrons and c for the neutral (now elastic) conglomerate. 
In order to avoid repetition, we use  the notation for the inner crust in the following equations but they are, when not explicitly stated, still valid 
 for the core once we  replace f with n and c with p. 

As we are working in Newtonian gravity, we cannot build our model from a realistic matter equation of state but we can still ensure that the phenomenology is appropriately represented. 
A simple model equation of state for a superfluid star is described 
by an energy functional which  ensures the Galilean invariance 
\citep{Prix}:
\begin{equation}
\mathcal{E} = \mathcal{E}_{0} \left(\rho_\x,  w_{\f \ch}^2 \right) \ , \label{eq:EoS}
\end{equation}
where $w_{\f \ch}^i = v_{\f}^i - v_{\ch}^i$ is the relative velocity between free neutrons and the neutral conglomerate. This leads to the chemical potential $\tilde \mu_\x $ and the entrainment parameter $\veps _\x $: 
\begin{equation}
\tilde \mu_\x \equiv  \left. \frac{ \partial \mathcal{E} }{ \partial \rho_\x }  \right| _{\rho_\y ,  w_{\x \y}^2} \label{eq:defmu}
\end{equation}
and 
\begin{equation}
2\alpha = \rho_\x \veps_\x \equiv  \left.  2 \frac{ \partial \mathcal{E} }{ \partial w_{\x \y}^2 }  \right| _{\rho_\x , \rho_\y  } \ . \label{eq:vareps}
\end{equation}
As the relative velocity between the two
fluids is expected to be small, equation~(\ref{eq:EoS}) can be expanded as:
\begin{equation}
\mathcal{E} = \mathcal{E}_{0} \left(\rho_\f, \rho_\ch \right)
+ \alpha_0 \left( \rho_\f, \rho_\ch \right) w_{\f \ch}^2 + \mathcal{O}\left(w_{\f \ch}^4\right) \ . \label{eq:EoSbulk}
\end{equation}
This approximation has the 
advantage that the bulk equation of state, $\mathcal{E}_{0}$, and the
entrainment parameter, $\alpha_0$, can be independently specified at
 $w_{\f\ch}^i=0$.  
From Eq.~(\ref{eq:vareps}) it follows
that the entrainment parameter $\varepsilon _{\x}$ is related to the
function $\alpha_0$ by
\begin{equation}
\rho_\x  \varepsilon_\x = 2 \alpha_0 \, .  \label{eq:alp}
\end{equation}
From the equation of state, we can derive a useful relation between density and chemical potential perturbations
\begin{eqnarray}
\delta \rho_\f  & = & \mathcal{S}_{\f\f}  \delta \tilde \mu _\f + \mathcal{S}_{\f\ch}  \delta \tilde \mu _\ch \, , \\
\delta \rho_\ch & = & \mathcal{S}_{\ch\f} \delta \tilde \mu _\f + \mathcal{S}_{\ch\ch} \delta \tilde \mu _\ch \, ,
\end{eqnarray}
which is valid for a co-moving background. Here we have defined
\begin{equation}
\mathcal{S}_{\x\y} \equiv \frac{\partial \rho_\x}{\partial \tilde \mu_\y} \, . \label{eq:Sdef}
\end{equation}

We build a simple model which allows for composition gradients, by combining two polytropes (as in \citealt{pass})
\begin{equation}
\mathcal{E}_{0} = k_\f \rho_\f ^{\gamma_\f} + k_\ch \rho_\ch
^{\gamma_\ch} \, , \label{eq:EosPR}
\end{equation}
where $k_\x$ and $\gamma_\x$ are constants. This equation of state  leads to models with varying composition
when $\gamma_\f \ne \gamma_\ch$.
From Eqs.~(\ref{eq:defmu})
and~(\ref{eq:EosPR}) it follows that the chemical potential and the corresponding mass density are related by
\begin{equation}
\rho_\x = \left( \frac{\tilde \mu _\x }{k_\x \gamma_\x } \right)
^{N_\x} \, ,   \label{eq:rhomu}
\end{equation}
where the polytropic index is given by $N_\x = \left( \gamma_{\x} -1
\right) ^{-1}$. From this result we can determine the proton fraction for a given stellar
 model by imposing  $\beta$-equilibrium. Assuming that the background star is in chemical equilibrium, and introducing  $\tilde \mu = \tilde \mu_\n= \tilde \mu_\ch$, we have
\begin{equation}
x_\ch = \left[ 1 + \frac{ \left( \gamma_\ch k_\ch \right) ^{N_\ch}
}{\left( \gamma_\f k_\f \right) ^{N_\f}} \, \tilde \mu ^{N_\f-N_\ch}
\right]^{-1} \, .  \label{eq:xp}
\end{equation}
For $\tilde \mu \to 0$, it is clear that
$x_\ch$ vanishes for $N_\f < N_\ch$ and tends to unity when $N_\f >
N_\ch$. 

\begin{figure}
\begin{center}
\includegraphics[height=70mm]{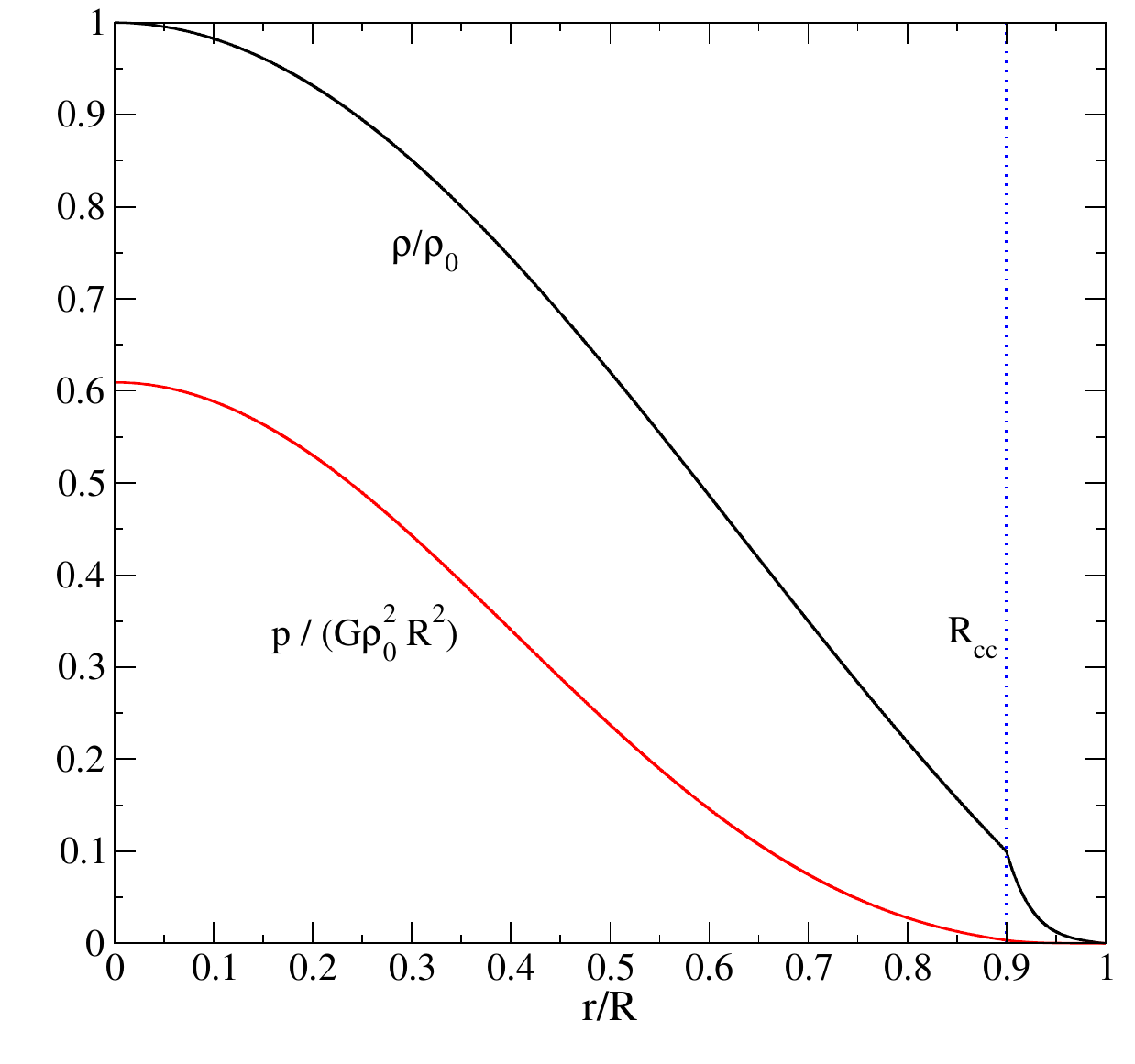}
\includegraphics[height=70mm]{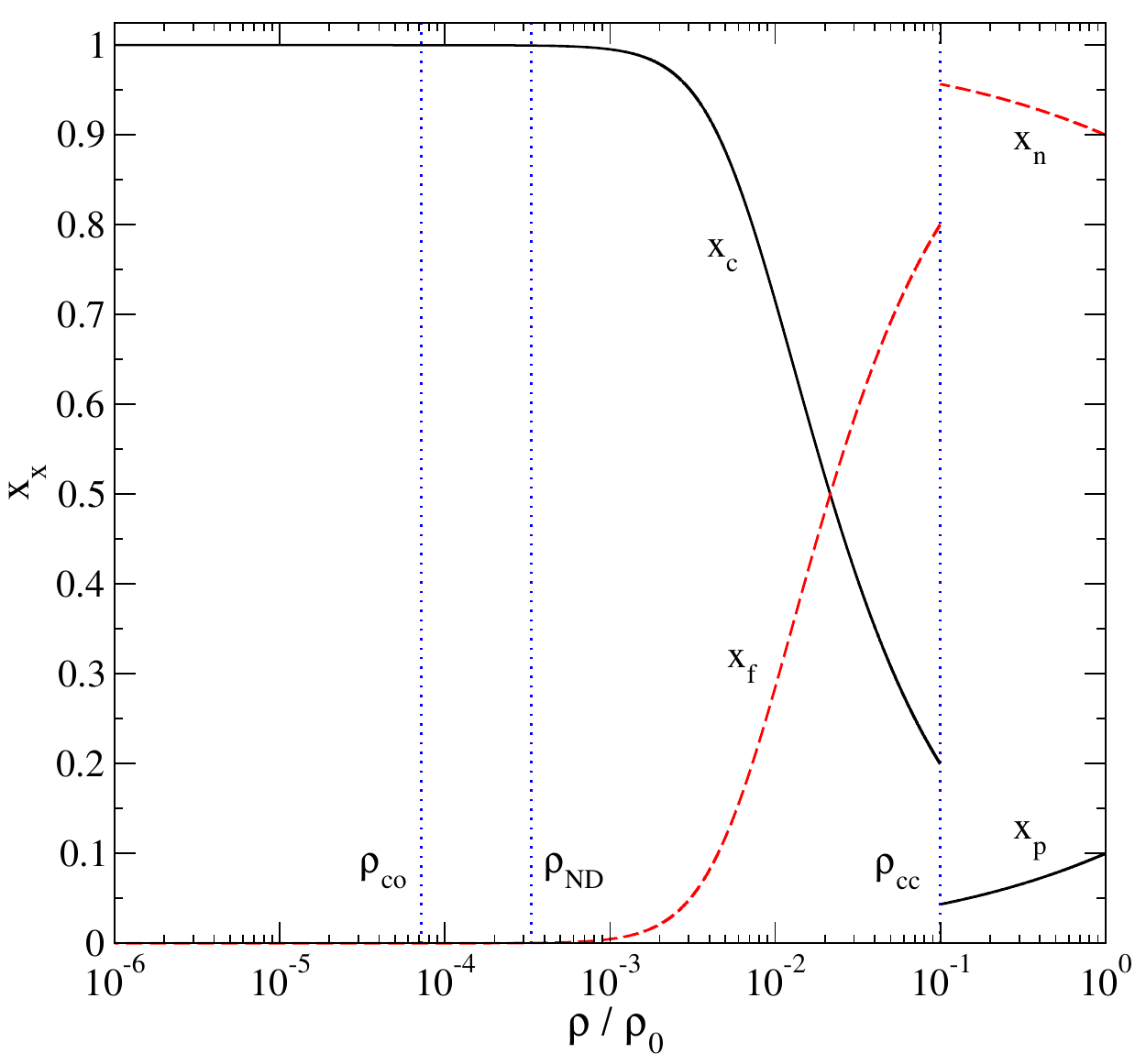}
\caption{ \label{fig:modelD}
Neutron star model with composition stratification. Top panel: radial profile of the total mass density and pressure, where $\rho_0$ is the central mass density. The vertical line denotes the location of the crust-core interface $R_{\ch\ch}$. Bottom panel: composition fractions 
with respect to the mass density. The transitions to the various layers in the star are indicated by mass densities $\rho_{\ch \ch}$, $\rho_{\rm ND}$, $\rho_{\ch o}$, representing the crust-core, neutron drip and crust-ocean transitions, respectively.}
\end{center}
\end{figure}

In realistic neutron star models, the proton fraction decreases from the centre to the crust-core interface, while the confined component
increases from the bottom of the crust and reaches unity at the neutron drip. We  approximate this behaviour by choosing in Eq.~(\ref{eq:EosPR}) 
different polytropic indices in the core and in the crust, $(N_\p, N_\n) \neq (N_\ch, N_\f)$. However, we assume that the total mass density,  the chemical potential and its derivative 
are continuous at the crust-core and  {crust-ocean} transition (see Fig.~\ref{fig:modelD}).

As a specific model we choose $N_\p=1.4$, $N_\n=1.0$, $k_\p=3.76$ and $k_\n=0.69$ in the core and 
$N_\ch=1.2$, $N_\f=3.9$, $k_\ch=1.88$ and $k_\f=0.20$ in the crust. 
The coefficients $k_\x$ are given in units of $G R^2 \rho_0^{2-\gamma_\x}$, where $\rho_0$ is the total mass density at the star centre. 
We set the crust-core transition at $R_{\ch\ch}=0.9 R$ and the crust-ocean interface at $R_{\ch \rm o}=0.999 R$, where $R$ is the star's radius.  
The solution of the equilibrium equations then leads to a star with mass $M_\star = 1.196 \, \rho_0 R^3$.  The proton fraction reaches 
 $x_\p  = 0.1$ at the centre and decreases to $4.3\times 10^{-2}$ at $R_{\ch\ch}$. The charged component at the bottom of the crust is  $x_\ch  = 0.2$ 
and increases to 1 at neutron drip.  The mass density, pressure and composition fractions are shown in Fig.~\ref{fig:modelD}. 

\begin{figure}
\begin{center}
\includegraphics[height=70mm]{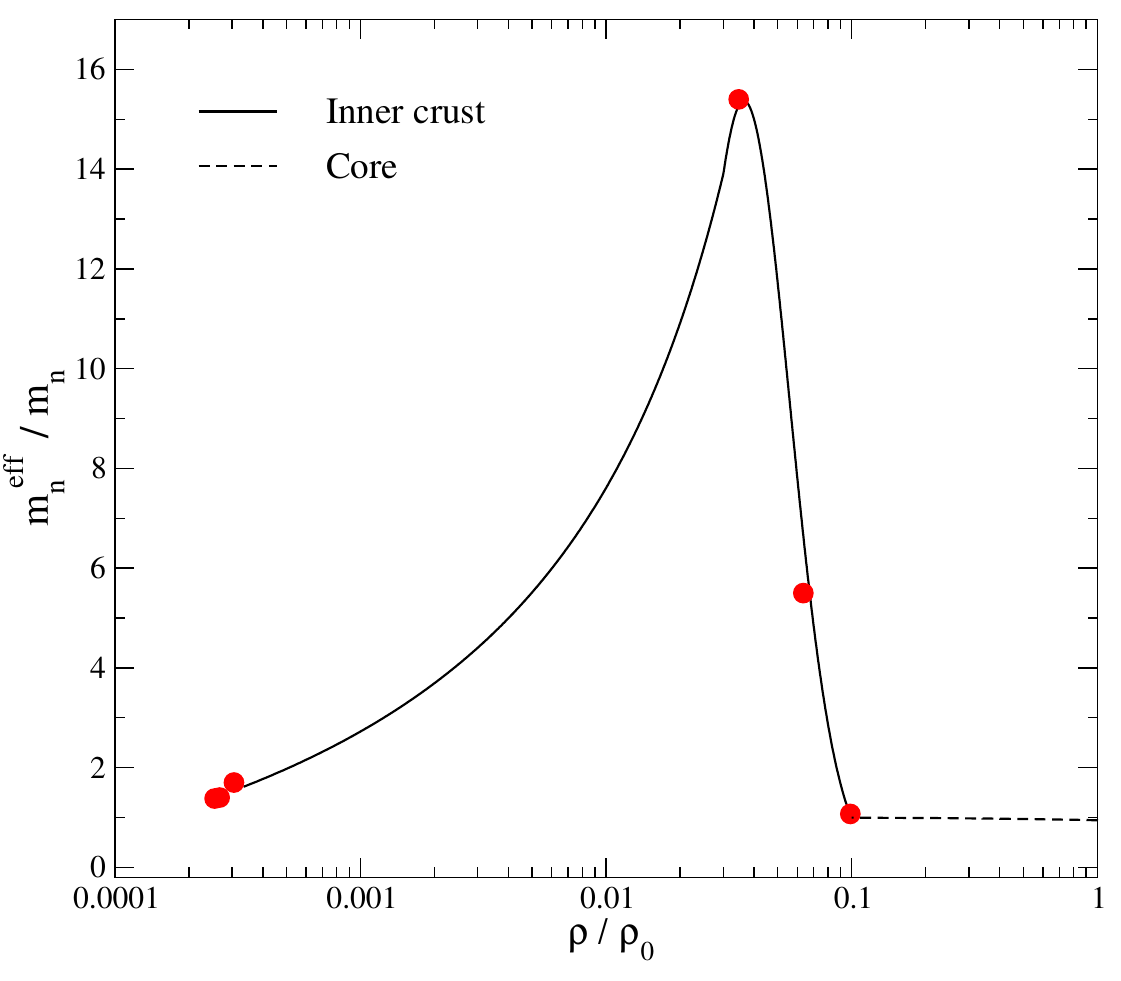}
\caption{ \label{fig:entr}
Effective neutron mass in the core and the inner crust. 
The red points indicate specific values taken from \citet{Carter05} and \citet{Ch05,Ch06}.}
\end{center}
\end{figure}

It is obviously helpful to express our solution in physical units in order to make contact with more `realistic' neutron star models. 
Considering that the  mass density at the crust-core transition is expected to be 
 $\rho_{\rm cc} = 1.2845 \times 10^{14} \, \textrm{g cm}^{-3}$  \citep{douchin}, we can 
use our numerical solution $\rho_{\rm cc} = 9.979 \times 10^{-2} \rho_0$ to determine the central mass density of our model,  $\rho_{0} = 1.29 \times 10^{15} \, \textrm{g cm}^{-3}$. In our model, neutron drip (expected to occur at around $\rho_{\rm ND} =4.3 \times 10^{11} \, \textrm{g cm}^{-3}$)  is reached at 
$R_{\rm ND}= 0.997 R$. 
We introduce the crust-ocean transition at $R_{\rm co} = 0.999 R$ where the mass density is $\rho_{\rm co} = 9.3 \times 10^{10} \, \textrm{g cm}^{-3}$. 
This value is larger than the typical melting point of a neutron star crust, $10^{6} - 10^{9} \, \textrm{g cm}^{-3}$. 
However, this is not a problem as long as we keep in mind that our results are  qualitative. More realistic neutron star models require relativistic gravity and tabulated equations of state.  
Finally, we can determine the stellar radius for a couple of typical masses. 
For $M_\star=1.4 M_{\odot}$ we have $R=12.2$~km, while for $M_\star=2.0~M_{\odot}$ the radius is $R=13.7$~km. 

In real superfluid neutron stars, the different constituents are coupled by different processes (entrainment, mutual friction, through the equation of state, etc.).
The entrainment, in particular,
can be described in terms of an effective mass  for each fluid component. 
The effect can be particularly strong in the inner crust 
\citep{Carter05,Ch05,Ch06}. We consider two indicative  models, one without entrainment and a second model with entrainment which has a large effective mass in the inner crust. 
Specifically, we use the same approximation as \citet{pass}. This leads to the effective neutron mass 
shown in Fig.~\ref{fig:entr}.

\begin{figure}
\begin{center}
\includegraphics[height=70mm]{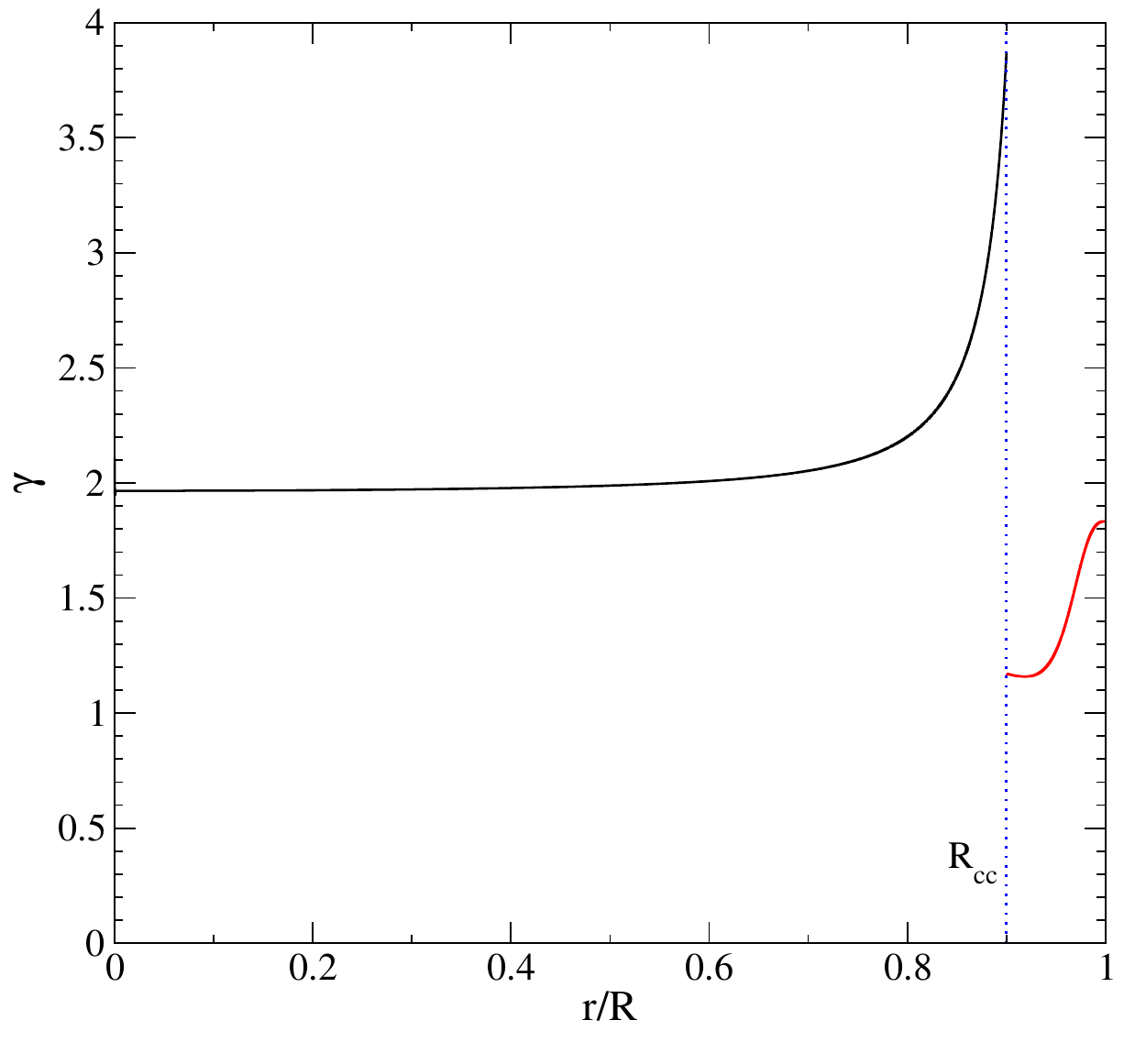}
\caption{ \label{fig:gammaD}
Radial profile of the polytropic index $\gamma$, obtained from Eq.~(\ref{eq:gam}), for the single-fluid reference model.}
\end{center}
\end{figure}

Finally, in order to establish the impact of superfluidity on the Love number we construct a single-fluid reference model with the 
main properties of the superfluid star. In order to do this,  we  need to provide the total mass density,  the pressure and the gravitational potential. In practice, all relevant variables follow from the inferred 
 profile for the polytropic index 
\begin{equation}
\gamma= \frac{\nabla \log p}{\nabla \log \rho} \, . \label{eq:gam}
\end{equation}
The variation of $\gamma$ with the radius is shown in Fig.~\ref{fig:gammaD}. 

\section{Results}

Having outlined the relevant theory and described the specific neutron star model, let us turn to the numerical results for
the mode contribution to the tidal response of a superfluid neutron star. 
As shown in Section~\ref{sec:sfproblem}, the oscillation modes of the homogeneous problem 
provide a complete set of eigenfunctions which can be used to decompose the tidal deformation during binary inspiral.  
In particular, some of the modes may become resonant with the tidal driving and, as a consequence, reach amplitudes such that they impact on the gravitational-wave signal or build up significant stresses in the crust.

We work with the perturbation equations from \citet{pass}, where the Lagrangian displacement for polar perturbations has the following spherical harmonic expansion (in the orthonormal basis used in the numerical code): 
\begin{align}
& \xi_r^\x = W_\x (r) Y_{lm} e^{i \omega t} \, , \\
& \xi_{\theta}^\x = V_\x (r) \partial _{\theta }Y_{lm} e^{i \omega t} \, , \\
& \xi_{\phi}^\x = V_\x (r) \frac{\partial _{\phi } Y_{lm}}{\sin \theta } e^{i \omega t} \, .
\end{align}

Before discussing the dynamical aspects, we consider the static limit and the impact of superfluidity on the 
Love number (the tidal deformability).  In the limit $m \tilde \Omega \ll \tilde \omega $, Eq.~(\ref{modesum}) is approximately given by 
\be 
k_{l} =  \sum_n  k_{n l} 
\label{modesum_st}
\ee
where the contribution to the  Love number from each mode is  given by
\be 
k_{nl} =   {2\pi \over 2l+1}  {|\tilde I_n|^2 \over  \tilde \omega_n^2 }  \, .
\label{knl_eff}
\ee
It is useful to note that in \citet{ap20a} and  \citet{PAP21} we  decomposed the Love number as 
\be 
k_{l} = - \frac{1}{2} + \sum_n  \bar k_{nl}
\label{modesum_st_old}
\ee
where $\bar k_{nl}$ can be written in several different, but equivalent ways \citep[see][]{ap20a}. One alternative is, for instance,
\be 
\bar k_{nl} =  {2\pi \over 2l+1} {|\tilde I_n|^2 \over  \tilde \omega_n^2 } \left[  1 -  \tilde \omega_n^2 \left( \frac{V_n}{W_n}\right)_{R} \right]^{-1} \, .
\label{knl_old}
\ee
The upshot is that the results obtained from (\ref{knl_eff}) for a single mode cannot be directly compared with those of our previous work, where we used equation (\ref{knl_old}). However, the combined results for the Love number are consistent. We have
\be 
\sum_n   k_{nl} = - \frac{1}{2} + \sum_n  \bar k_{nl} \, , 
\label{cons}
\ee
as a direct consequence of equations (\ref{modesum_st}) and (\ref{modesum_st_old}), and we have confirmed that the numerical results satisfy this relation.

\begin{table}
\begin{center}
\caption{Mode results for the neutron star model without entrainment. The first four columns,
respectively, report the oscillation mode, the dimensionless mode frequency, the absolute value
of the dimensionless integral $\tilde I_n$, and the ratio between the radial and tangential components
of the displacement vector at the surface. The last column provides the contribution each mode 
makes to the static Love number, $k_l$. The final row shows the summed result.
\label{tab:tab1} }
\begin{tabular}{c c c c c  }
\hline
 $ \rm Mode $ &  $ \tilde \omega_n $  &  $ | \tilde I_n | $   & $ \left( V/W \right)_R $  & $ k_{nl}$  \\ 
\hline
p$_3$  & 6.8370	  &  $3.0170 \times 10^{-4}$  &  $2.1393\times 10^{-2}$ & $2.45 \times 10^{-9}$ \\  
p$_2$  & 4.8076	  &  $1.1169 \times 10^{-3}$  &  $4.3264\times 10^{-2}$ & $6.78\times 10^{-8}$ \\ 
p$_1$  & 3.5840	  &  $2.0354 \times 10^{-2}$  &  $7.7946\times 10^{-2}$ &  $4.05\times 10^{-5}$ \\ 
f$ _{\hspace{-0.05mm}\rm o}$      & 1.2693	  &  $5.4579 \times 10^{-1}$  &  $4.2829\times 10^{-1}$  & $0.23234$ \\ 
\\
p$_3 ^{\hspace{-0.05mm}\s}$& 7.0777 & $	2.4274\times 10^{-4}$ &  $1.9961 \times 10^{-2}$  & $1.48\times 10^{-9}$ \\ 
p$_2 ^{\hspace{-0.05mm}\s}$& 5.8051 & $	1.4477\times 10^{-3}$ &  $2.9667 \times 10^{-2}$ & $7.82\times10^{-8}$ \\ 
p$_1 ^{\hspace{-0.05mm}\s}$& 4.1830 & $	4.8759\times 10^{-4}$ &  $5.7098 \times 10^{-2}$ & $1.71\times 10^{-8} $ \\ 
f$ _{\hspace{-0.05mm}\s}  $& 2.0449 & $	4.1092\times 10^{-3}$ &  $2.3869 \times 10^{-1}$ & $5.07\times 10^{-6}$ \\  \\ 
s$_1$  & 0.4612 & $4.3945 \times 10^{-4}$  &  $4.6786$ & $1.14\times 10^{-6}$ \\ 
s$_2$  & 0.8391 & $6.8147 \times 10^{-4}$  &  $1.4351$ & $8.29\times 10^{-7}$\\ 
s$_3$  & 1.2118 & $6.6749 \times 10^{-3}$  &  $0.6500$ & $3.81\times10^{-5}$ \\ 
s$_4$  & 1.5834 & $9.4790 \times 10^{-4}$  &  $0.3943$ & $4.50\times 10^{-7}$\\ 
s$_5$  & 1.9546 & $4.6317 \times 10^{-4}$  &  $0.2625$ & $7.06\times 10^{-8} $ \\ 
s$_6$  & 2.3258 & $2.7882 \times 10^{-4}$  &  $0.1843$ & $1.81\times 10^{-8}$ \\
\\ 
 i$_2$    &0.0564 & $8.1268 \times 10^{-5}$  &  $3.1421 \times 10^{2}$ &  $2.61\times 10^{-6}$ \\ 
 i$_1$    &0.0134 & $5.8098 \times 10^{-6}$  &  $5.5648 \times 10^{3}$ & $2.36\times10^{-7}$ \\ 
 \hline 
 $k_l$ & & & & 0.23242  \\
 \hline   
\end{tabular}
\end{center}
\end{table}

\begin{table}
\begin{center}
\caption{Same as Table \ref{tab:tab1} for the model with entrainment. The s$_4$-mode line is left blank because this mode lies in the same frequency range with the ordinary f-mode, therefore numerical convergence is difficult to achieve.
\label{tab:tab2} }
\begin{tabular}{c c c c c  }
\hline
 $ \rm Mode $ &  $ \tilde \omega_n $  &  $| \tilde I_n | $   & $ \left( V/W \right)_R $  & $ k_{nl}$ \\ 
\hline
p$_3$    &  6.9820	&   $8.0756 \times 10^{-6}$  &  $2.0513 \times 10^{-2}$ & $1.68\times10^{-12}$\\ 
p$_2$    &  5.5401	&   $1.3037 \times 10^{-3}$  &  $3.2580 \times 10^{-2}$ & $6.96\times10^{-8}$\\ 
p$_1$    &  3.7275	&   $1.8984 \times 10^{-2}$  &  $7.2111 \times 10^{-2}$ & $3.26\times10^{-5}$  \\  
f$ _{\hspace{-0.05mm}\rm o}$        &  1.2695	&   $5.4565 \times 10^{-1}$  &  $4.1420 \times 10^{-1}$ & $0.23217$ \\ \\ 
 p$_2 ^{\hspace{-0.05mm}\s}$  & 2.6996 &  $6.6633 \times 10^{-3}$ & $1.3641 \times 10^{-1}$ & $7.66\times 10^{-6}$\\ 
 p$_1 ^{\hspace{-0.05mm}\s}$  & 2.5083 &  $8.9742 \times 10^{-4}$ & $1.5903 \times 10^{-1}$ & $1.61\times10^{-7}$ \\ 
 f$ _{\hspace{-0.05mm}\s}$    & 1.0265 &  $1.0173 \times 10^{-2}$ & $9.4643 \times 10^{-1}$ & $1.23\times10^{-4}$
\\ \\ 
s$_1$    & 0.3836  & $5.1906\times 10^{-4}$  &  $6.7766$ & $2.30\times10^{-6}$ \\ 
s$_2$    & 0.6906  & $4.7713\times 10^{-4}$  &  $2.1180$ & $6.00\times10^{-7}$ \\ 
s$_3$    & 0.9898  & $3.7483\times 10^{-4}$  &  $1.0200$ & $1.80\times10^{-7}$\\ 
s$_4$    &   &  \\ %
s$_5$    & 1.5898  & $7.8292\times 10^{-4}$  &  $0.3986$ & $3.05 \times 10^{-7}$ \\ 
s$_6$    & 1.8891  & $3.0885\times 10^{-4}$  &  $0.2793$ &  
$3.36\times 10^{-8}$ \\ \\ 
 i$_2$    & 0.0376  &	  $5.8390 \times 10^{-5}$  &  $7.0752 \times 10^{2}$  & $3.03\times 10^{-6}$ \\    
 i$_1$    & 0.0134  &	  $5.9756 \times 10^{-6}$  &  $5.5659 \times 10^{3}$ & $2.50\times 10^{-7}$\\    
 \hline 
$k_l$ & & & &  0.23234  \\
 \hline   
\end{tabular}
\end{center}
\end{table}
\begin{table}
\begin{center}
\caption{Same as Table \ref{tab:tab1} for the barotropic  single-fluid model. \label{tab:tab3} }
\begin{tabular}{c c c c c  }
\hline
 $ \rm Mode $ &  $ \tilde \omega_n $  &  $| \tilde I_n |$   & $ \left( V/W \right)_R $  & $ k_{nl}$ \\ 
\hline
 p$_3$    &  6.7041 &   $5.6000 \times 10^{-4}$  &  $2.2249 \times 10^{-2}$   & $8.77\times10^{-9}$ \\ 
 p$_2$    &  5.1425 &   $3.0948 \times 10^{-3}$  &  $3.7810 \times 10^{-2}$   & $4.55\times 10^{-7}$\\ 
 p$_1$    &  3.7318 &	$2.3154 \times 10^{-2}$  &  $7.1951 \times 10^{-2}$   & $4.84\times 10^{-5}$  \\ 
 f        &  1.2693 &   $5.4582 \times 10^{-1}$  &  $4.2845 \times 10^{-1}$   & 0.23237 \\ \\
 
 s$_1$    & 0.3766 & $4.0011 \times 10^{-4}$  &  $7.0033$   & $1.42\times 10^{-6}$ \\ 
 s$_2$    & 0.6598 & $3.7305 \times 10^{-4}$  &  $2.3103$   & $4.02\times 10^{-7}$ \\ 
 s$_3$    & 0.9471 & $9.0455 \times 10^{-4}$  &  $1.1029$   & $1.15\times 10^{-6}$ \\  
 s$_4$    & 1.2346 & $5.4510 \times 10^{-3}$  &  $0.7079$   & $2.45 \times 10^{-5}$ \\ 
 s$_5$    & 1.5223 & $9.2691 \times 10^{-4}$  &  $0.4357$   & $4.66 \times 10^{-7}$ \\ 
 s$_6$    & 1.8100 & $3.1379 \times 10^{-4}$  &  $0.3043$   & $3.78 \times 10^{-8} $ \\ 
\\ %
 i$_2$    & 0.0268 & $7.8216 \times 10^{-5}$  &  $1.3878 \times 10^{3}$   & $1.07\times 10^{-5} $ \\  
 i$_1$    & 0.0134 & $4.8537 \times 10^{-6}$  &  $5.5665 \times 10^{3}$   & $1.65\times 10^{-7}$ \\ 
 \hline 
$k_l$ & & & &  0.23246 \\ 
 \hline   
\end{tabular}
\end{center}
\end{table}
In a superfluid star the mode classification is richer than for single fluid models. The presence of the additional degree of freedom---the superfluid neutrons---doubles the family of the fundamental and acoustic modes \citep{epstein, mend1,mend2, Lee, AC01,PR02}. As in a coupled harmonic oscillator, the two-fluid oscillations can be  described  
 by  co- and counter-moving motion.  The first  leads to oscillation modes with properties very similar to the 
 single fluid model, often called `ordinary modes'. The counter-moving oscillations instead occur only in superfluid stars and for this reason tend to be referred to as `superfluid modes'. 
 With our model we can study the ordinary and superfluid fundamental modes (f-modes) and pressure modes (p-modes). 
In addition, as in \citet{PAP21}, the presence of an elastic crust sustains shear modes (s-modes), while any sharp transition between  different stellar regions produces an interface mode (i-modes).  
We cannot study here the gravity modes (g-modes), which are restored by buoyancy, because in superfluid stars these appear only
when there is at least a third constituent, as for instance muons (see \citealt{GK13}, \citealt{KG14} and \citealt{PAH16}). These modes were, however, discussed by \citet{wein}, so it makes sense for us to focus on the two-fluid aspects of the problem.
\begin{figure*}
\begin{center}
\includegraphics[height=70mm]{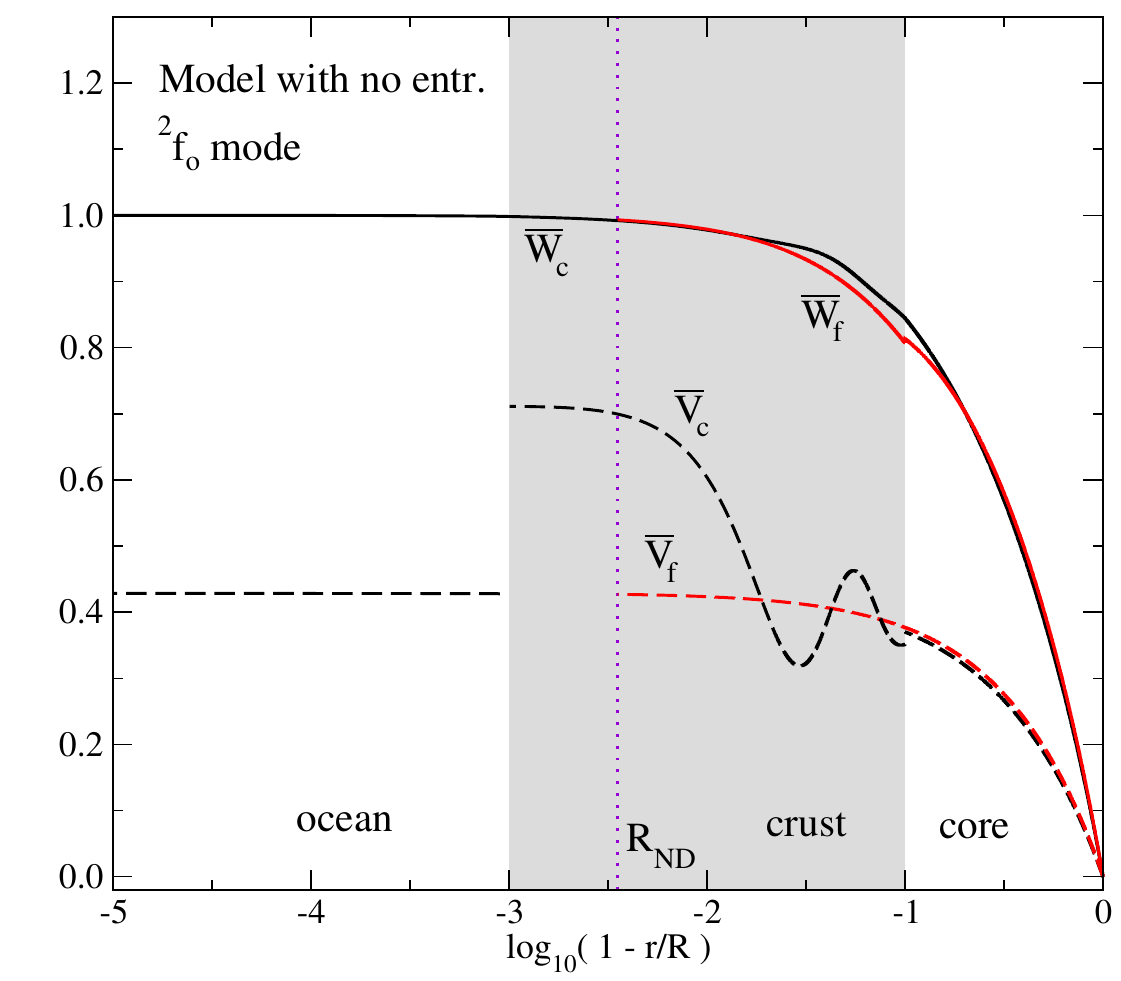}
\includegraphics[height=70mm]{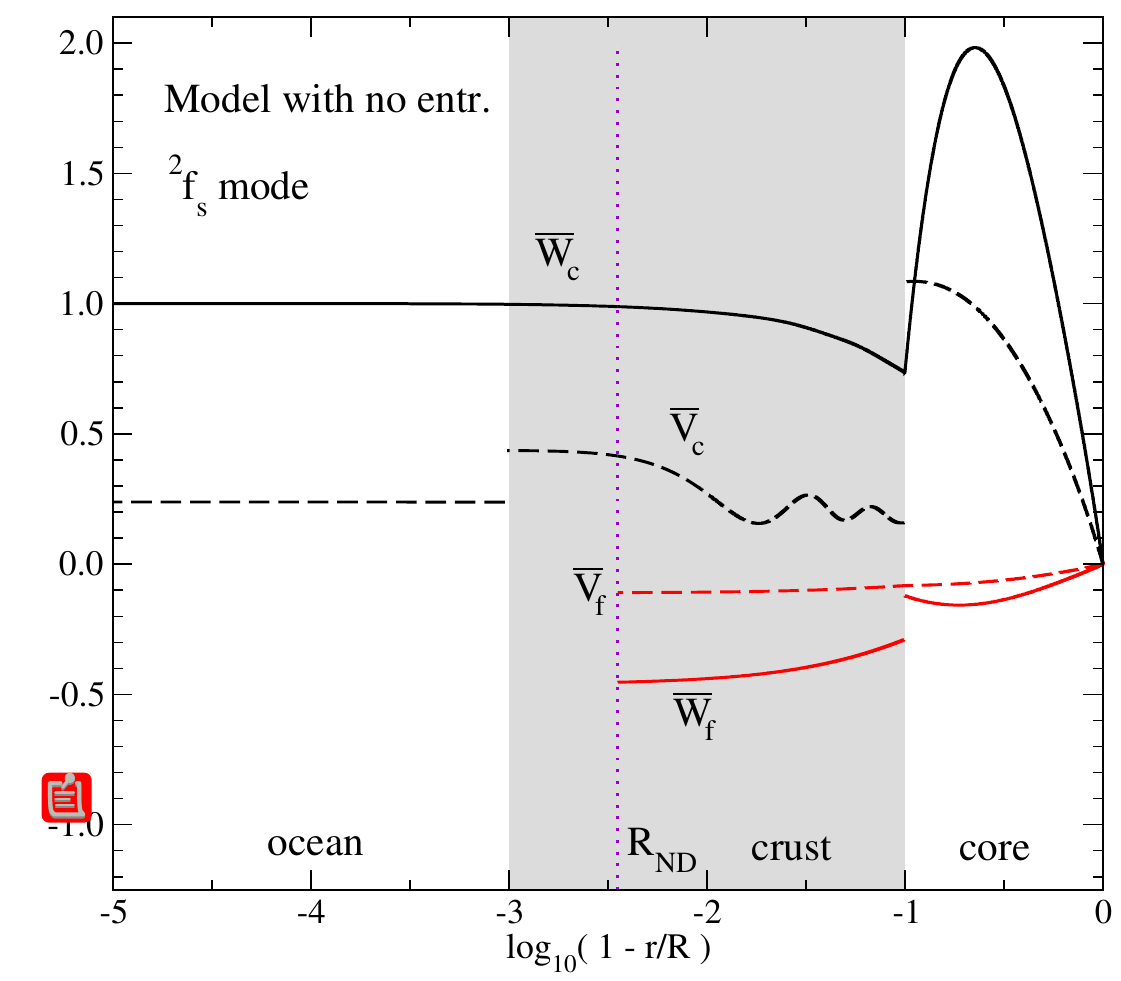}
\caption{ \label{fig:eig-f}
Eigenfunctions of the Lagrangian displacement for the ordinary and superfluid f-modes, for the model without entrainment. 
The vertical dotted line indicates the location of the neutron drip transition $R_{\rm ND}$.
The left panel shows the ordinary f-mode while the right panel depicts the superfluid f-mode. The solid and dashed lines show, respectively, 
the functions $\overline W_\x = W_\x/R$ and $\overline V_\x=V_\x/R$, where $R$ is the star's radius. Black lines refer to the neutral conglomerate of charged particles, while the red lines to the superfluid neutrons in the core and in the inner crust. \label{fig:eig1}
}
\end{center}
\end{figure*}
\begin{figure*}
\begin{center}
\includegraphics[height=70mm]{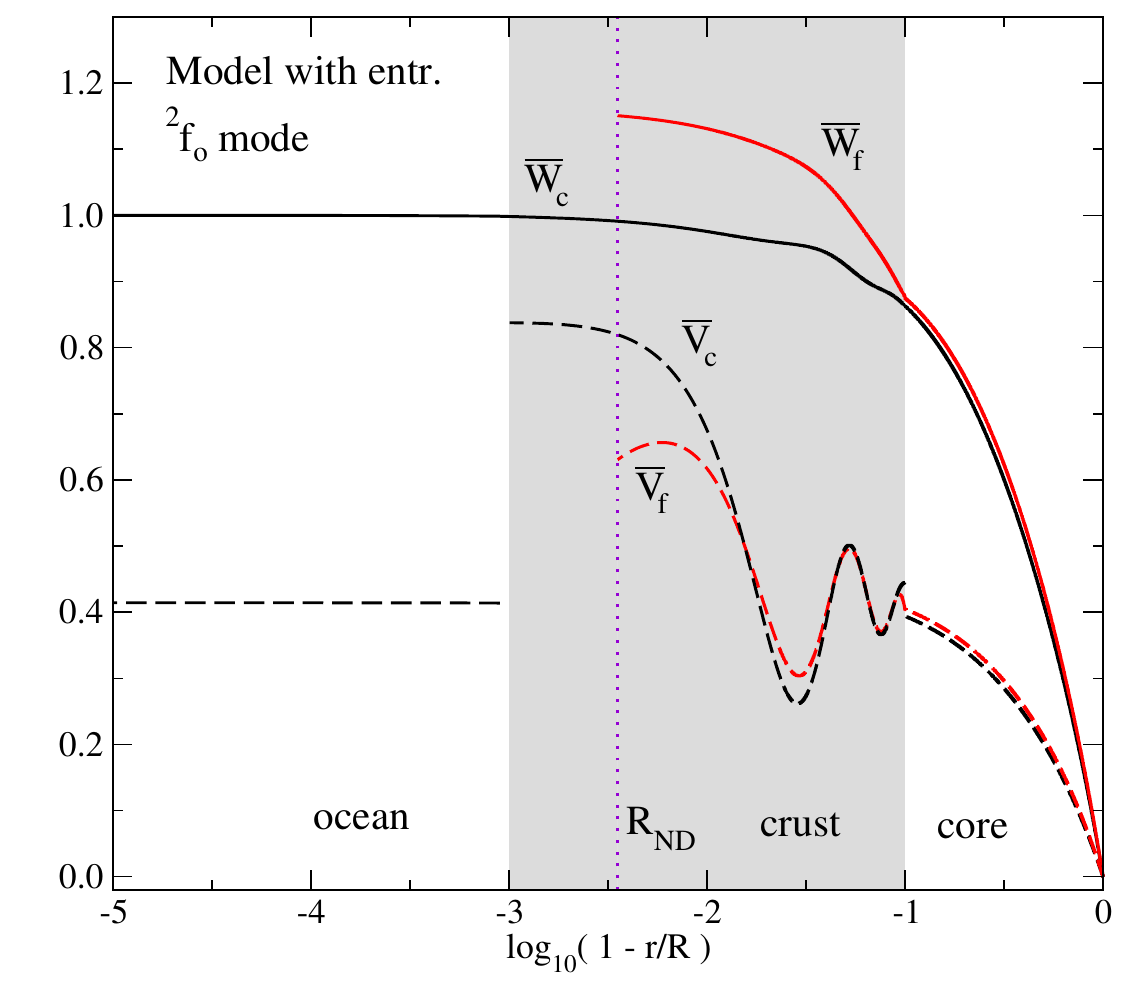}
\includegraphics[height=70mm]{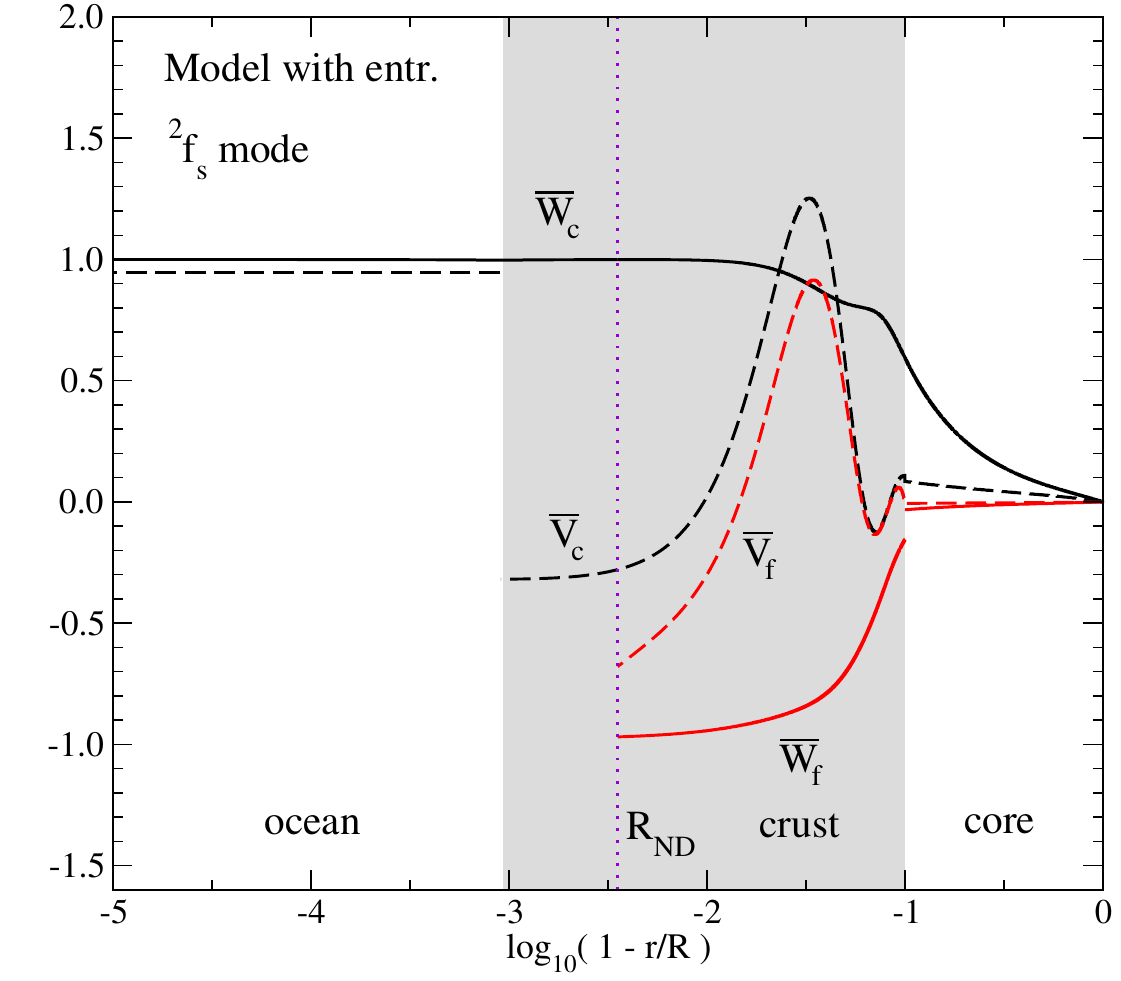}
\caption{ \label{fig:eig-f-entr}
Same as Fig.~\ref{fig:eig1}, but for a star with strong entrainment.}
\end{center}
\end{figure*}
The impact of  entrainment  can be significant on both mode frequencies and  eigenfunctions. 
In general, entrainment increases the coupling between the two components,  enforcing the comoving  features and 
leading to oscillation properties more similar to the single fluid case. The effect of entrainment on the various oscillation modes has already been extensively studied in literature \citep[see for instance][]{pass}.

We determine the mode properties by solving the linearised equations as an eigenvalue problem using  
multiple shooting and relaxation methods, updating the numerical code developed by \citet{PAP21} taking care of the 
junction conditions at the internal boundaries. The relaxation approach is used to increase the numerical accuracy of the solution, 
especially for calculating the required overlap integrals. The multiple shooting method provides a first guess solution for the relaxation code. 
More details on the numerical codes and relevant tests are provided by \citet{PAP21}.

For the superfluid model without entrainment the main properties of the oscillation modes are shown in Table \ref{tab:tab1}. 
We focus on the $l=2$ multipole which is the most relevant for the tidal deformation and the gravitational-wave aspects. As anticipated,
the ordinary f-mode has (by far) the largest overlap integral and provides the main contribution to the effective Love number. 
The shear and interface mode results are similar to the single-fluid model, and it is evident that their 
influence on the dynamical tide is negligible (at least away from resonance). 
The superfluid f- and p-modes are associated with very small values for $ \tilde I_n$  and $k_{nl}$. This is  expected, since the two fluid components are weakly coupled in this model. The superfluid modes are dominated by counter-moving motion and their contribution to the total mass density is small compared to the co-moving oscillations. Hence, they are not significantly affected by the tidal driving.
This behaviour is even more evident in the two-fluid model without crust. 
The eigenfunctions of the Lagrangian displacement for the ordinary and superfluid f-modes are shown in Fig.~\ref{fig:eig-f},  where $\overline{W}_\x = W_\x /R$ and $\overline{V}_\x = V_\x /R$ and $R$ is the star's radius. 
The results in the two panels clearly illustrate the superfluid nature of the $^2$f$_\s$-mode, where the superfluid and normal constituents tend to have eigenfunctions with opposite sign (they are counter-moving).  
\begin{figure*}
\begin{center}
\includegraphics[height=70mm]{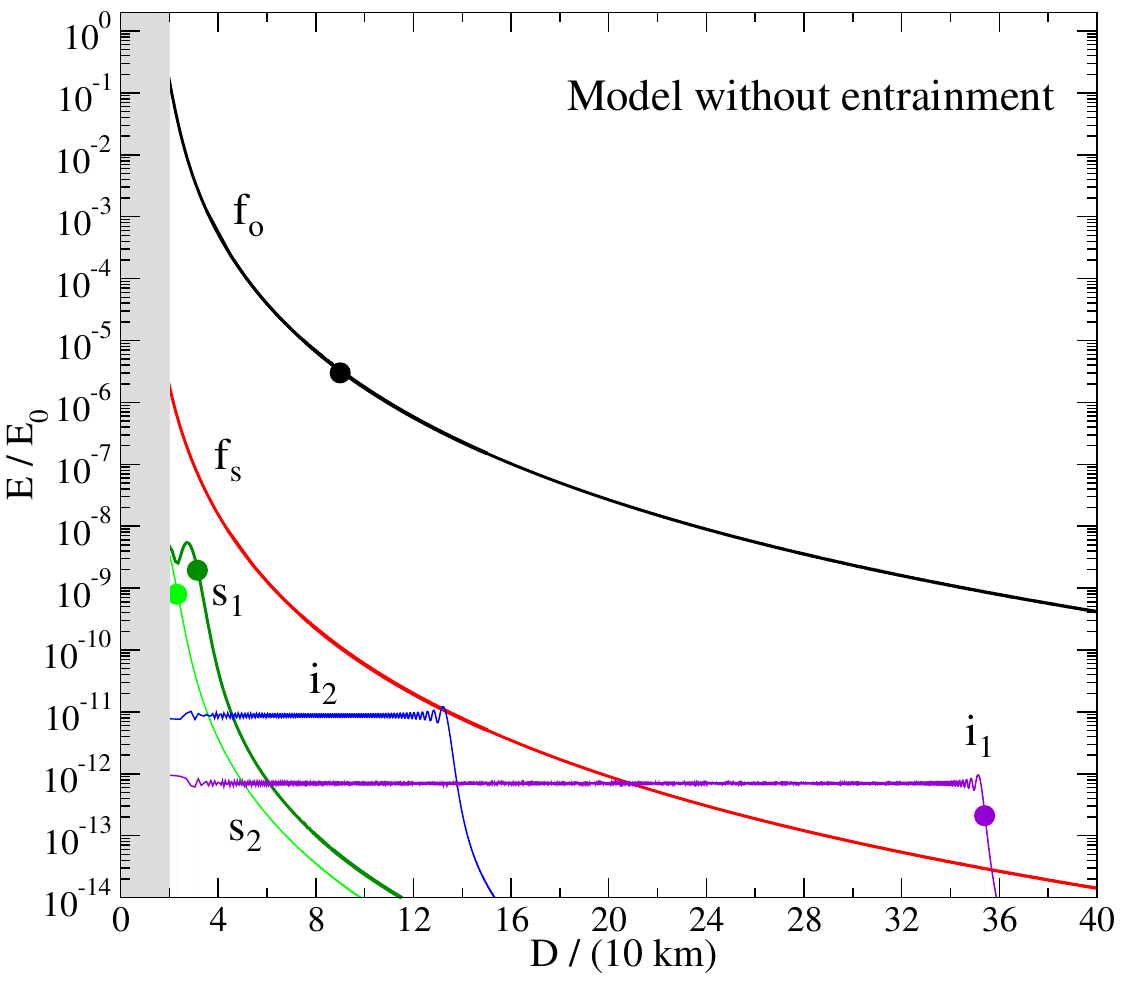}
\includegraphics[height=70mm]{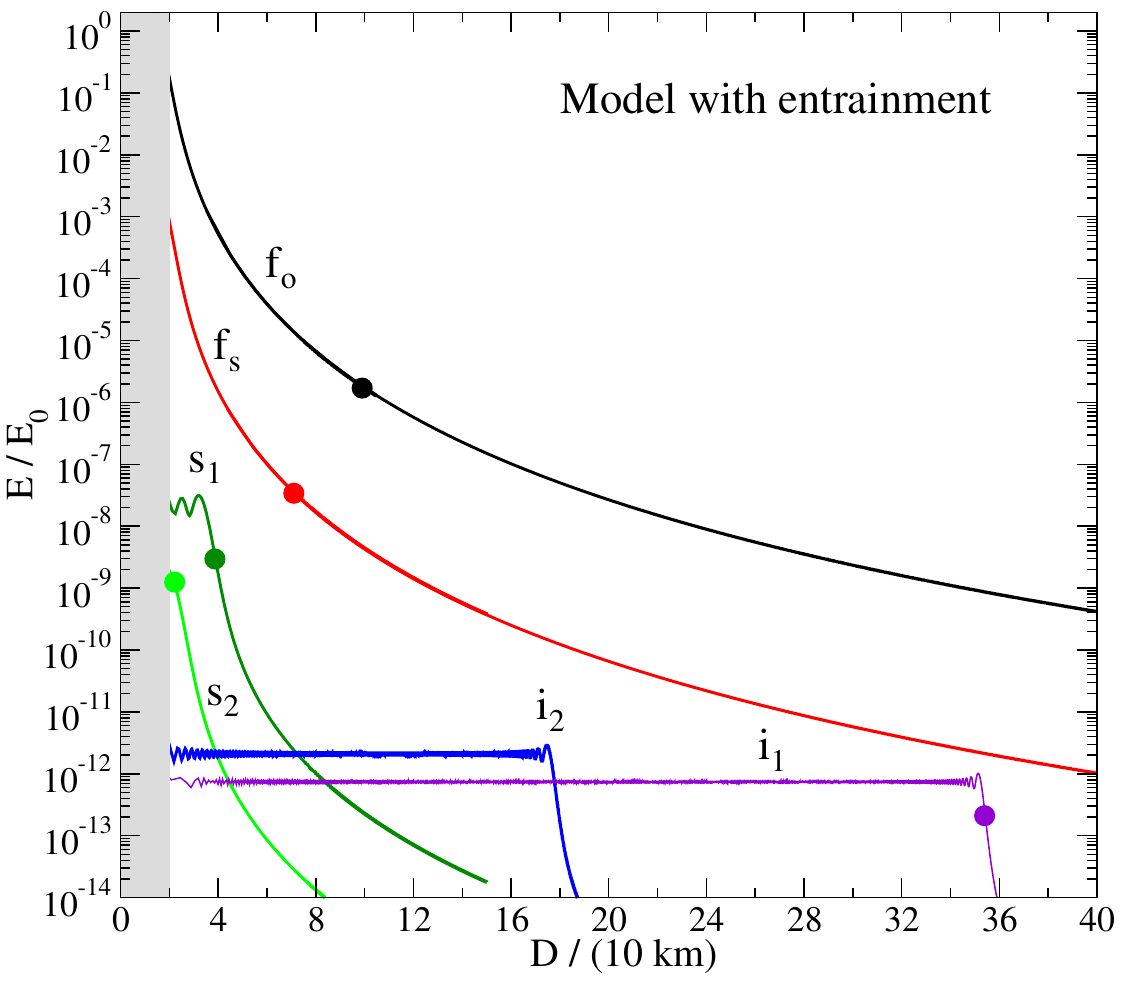}
\caption{ \label{fig:En1}
Total energy of the most relevant oscillation modes during the inspiral 
for the stellar models with entrainment (right panel) and without entrainment (left panel).  
On the horizontal axis we show the orbital separation, $D$, and on the vertical axis the total mode energy, normalised to $E_0 = G M_\star^2 / R $. The breaking energy limit $E_b$ of the crust is represented by a circle which corresponds to a breaking strain $\bar\sigma_b = 0.1 $.}
\end{center}
\end{figure*}
The figure also shows the discontinuity between the radial component  $\overline W_\f $ and 
$\overline W_\n$, which is an effect of the junction conditions  
between the superfluid neutrons of the core and inner crust (see the discussion in Section~\ref{sec:sfproblem}).

When we add entrainment to the model there are notable changes in the frequencies of the superfluid modes. 
The $^2$f$_\s$-mode changes from $\tilde \omega_n = 2.0449$ for the model without entrainment to $\tilde \omega_n = 1.0265$ for the specific entrainment model we consider. This latter value is lower than the ordinary fundamental mode frequency, $\tilde \omega_n = 1.2695$, which is hardly at all affected by entrainment. This variation is explained by the (assumed) large effective neutron mass in the inner crust, which strongly couples  the dynamics of the two constituents.  
From the eigenfunctions of the Lagrangian displacement for the ordinary and superfluid f-modes, shown in Fig.~\ref{fig:eig-f-entr}, we note that the two fluid constituents of the $^2$f$_\s$-mode maintain the counter-moving character in the 
radial components $\overline W_\ch$ and $\overline W_\f$, while the tangential component $\overline V_\f$ in the inner crust has the same sign 
as $\overline V_\ch$ and their radial profiles are very similar. Actually, at the bottom of the crust, they almost overlap. 
The same feature is exhibited by the tangential components of the $^2$f$_{\rm o}$-mode; compare Fig.~\ref{fig:eig-f-entr} with Fig.~\ref{fig:eig-f}. 
For the superfluid pressure modes, we find similar changes in the oscillation frequencies   (see Table \ref{tab:tab2}).

Not surprisingly, the ordinary f-mode provides the most relevant contribution to the Love number even for the model with entrainment. Its value, $k_{nl} = 0.23217$, is only slightly smaller than in the case without entrainment, $k_{nl} = 0.23234$. The difference, $1.7 \times 10^{-4}$ (much too small to ever be distinguished by observations), is mainly explained by the superfluid 
f-mode contribution, $k_{nl}=1.23\times10^{-4}$,  more than an order of magnitude larger than in the non-entrained case (see Tables \ref{tab:tab1} and \ref{tab:tab2}). The results for $k_{nl}$ for the superfluid pressure modes remain negligible even though they show a slight enhancement due to  entrainment. In essence,  the superfluid parameters affect the mode results but the overall conclusion remains the same. The tidal response is completely dominated by the  ordinary f-mode. 
Still,  our results quantify the extent to which entrainment  impacts on the Love number. 
The mode sum for the model without entrainment leads to $k_l = 0.23242$, which is only slightly larger than the result for the model with 
entrainment, $k_l = 0.23234$. However, the difference, about $8\times10^{-5}$, is small enough that it may be due to numerical error---we are certainly not claiming that the mode results are accurate to 5 decimal places. The main conclusion is that we do not distinguish an influence of entrainment on the Love number. This accords with the argument from Section~\ref{sec:sfproblem}, which explains why the result in the static limit should not be affected by superfluidity.

In order to further test this assertion,  we consider the results for a single fluid model, corresponding to a star with normal matter with the same pressure and total mass density as the superfluid model (see Section \ref{sec:NS}). In this case,
the mode sum provides $k_{l} = 0.23246$ and, not surprisingly, the fundamental mode gives the main 
contribution (see Table \ref{tab:tab3}). Again, the value is very close to the two superfluid models, in particular the model with entrainment. In essence, there is no (significant) difference between the static tide in superfluid and normal stars.   


\subsection{Mode excitation}

In addition to the mode-sum for the  {static} tide, we may consider the tidal excitation of the stellar oscillations, focusing on the most relevant modes. 

We can apply the strategy and formalism developed for single fluid stars \citep{PAP21}, the only difference being the normalisation $\mathcal{A}_n^2$ which is now given by Eq.~(\ref{entnorm}). 
The mode amplitude evolution is described by \citep{Lai}
\begin{equation}
\ddot a_n + \omega_n ^2 a_n = -\frac{G M' }{R^3}  
\tilde I_n  \left( \frac{R }{D(t)} \right)^{l+1}   W_{lm} \, e^{- i m \Phi(t)} \label{eq:amp}
\end{equation}
where $D$ is the orbital separation, $\Phi (t) = \int \Omega(t) dt$ and for $l=2$ the $W_{lm}$ coefficients are \citep{Lai}
\begin{equation}
W_{20} =  - \sqrt{\frac{\pi}{5}} \ , \quad\quad  W_{2\pm2} =   \sqrt{\frac{3 \pi}{10}} \ , \quad\quad  W_{2\pm1} =  0 \ .
\end{equation}
The evolution of the orbital separation and the orbital frequency follows from 
\begin{gather}
 \dot D  = - \frac{64}{5} \frac{G^3}{c^5} \frac{M_\star M' (M_\star+M')}{D^3} \, , \\
 \Omega = \left[  \frac{G  (M_\star+M')}{D^3}  \right]^{1/2} \, .
\end{gather}
From the evolution of the mode amplitude we can determine the kinetic and potential energy and therefore the 
total mode energy:
\begin{equation}
E = E_k  + E_p = \frac{1}{2} \sum_n \mathcal{A}_n^2 \, \left( |\dot a_n(t)|^2  + \omega_n^2 |a_n(t)|^2  \right) \, .
\end{equation}
More details on the numerical method used to evolve Eq.~\eqref{eq:amp}  are given by \citet{PAP21}.

\begin{figure}
\begin{center}
\includegraphics[height=70mm]{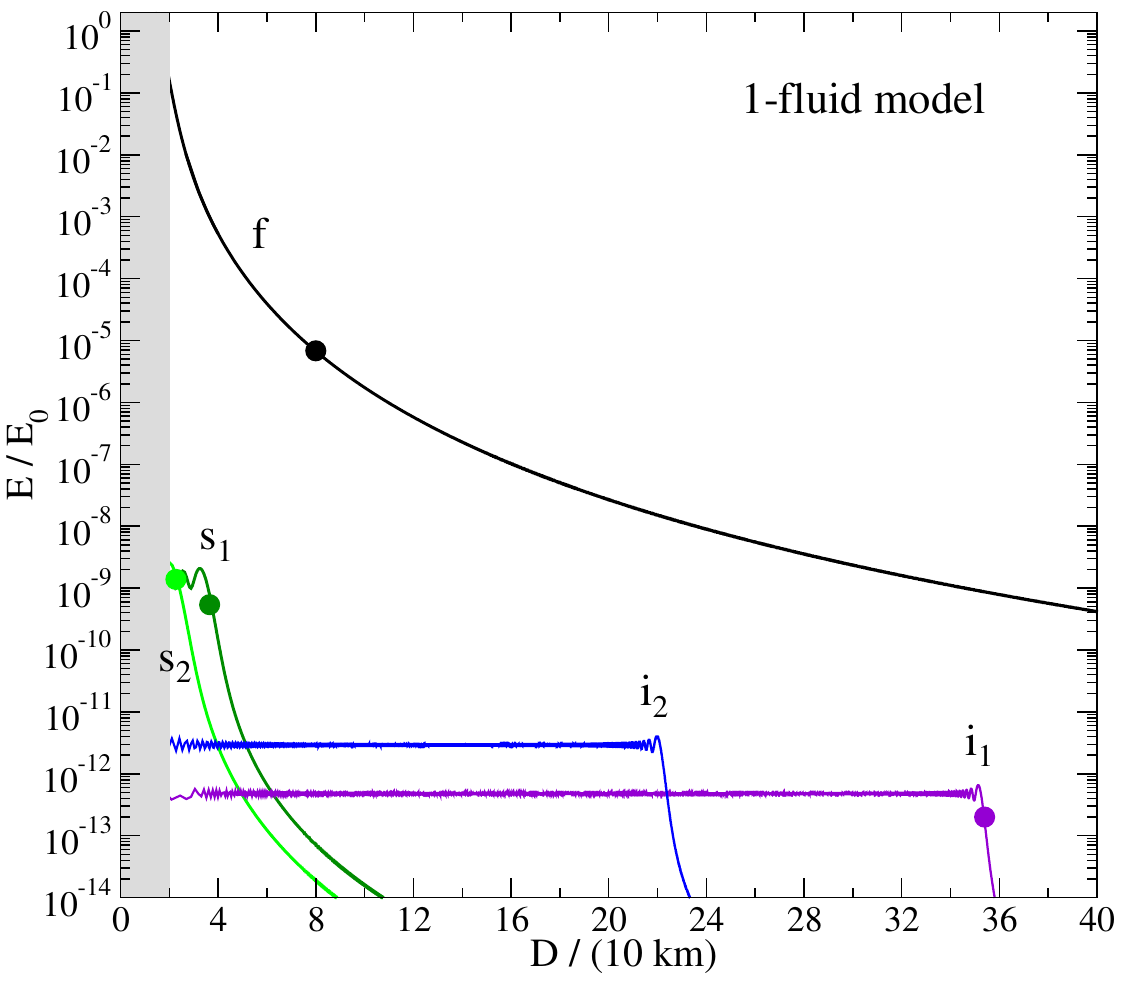}
\caption{ \label{fig:En2}
Same as Fig.~\ref{fig:En1}, but for the single-fluid model.}
\end{center}
\end{figure}

\begin{figure}
\begin{center}
\includegraphics[height=70mm]{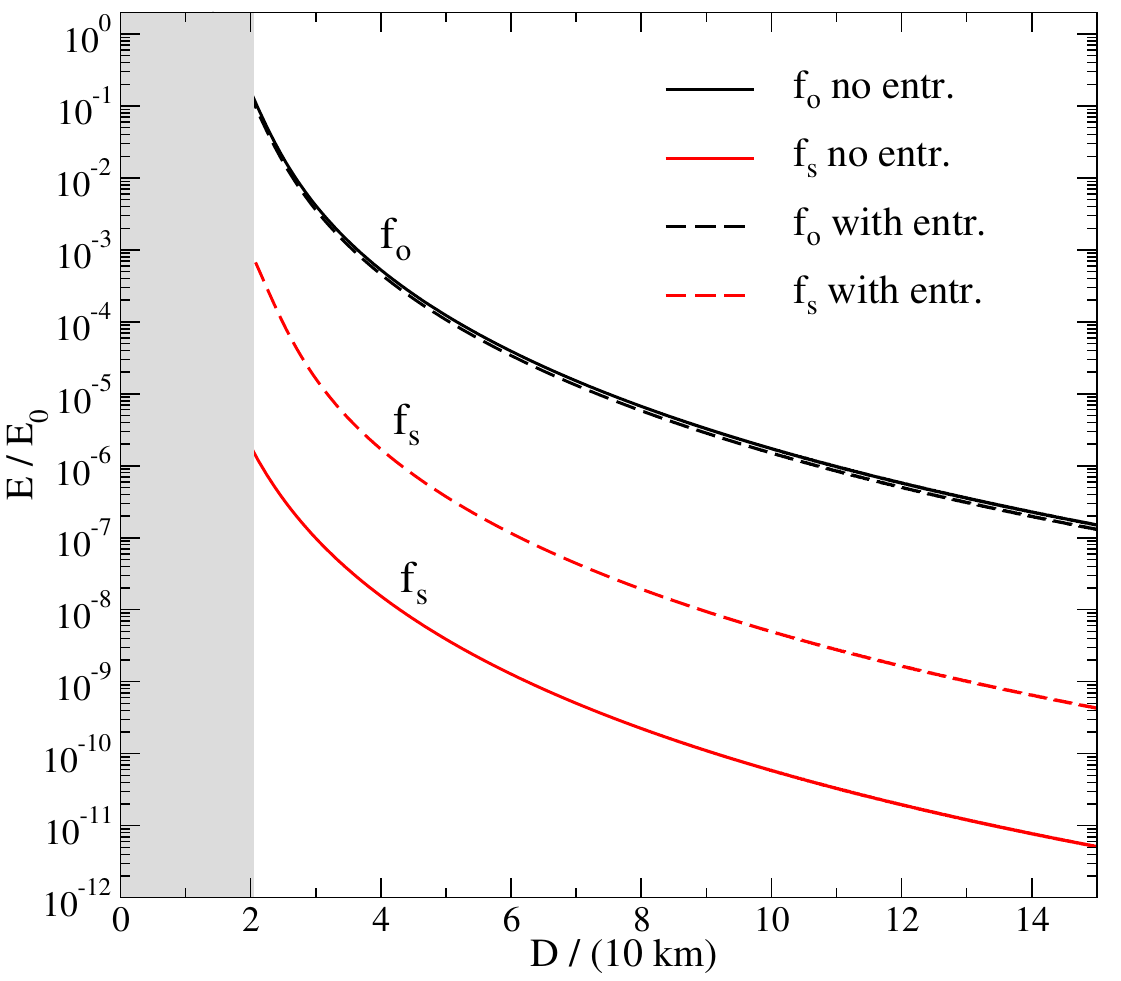}
\caption{ \label{fig:En3}
Energy evolution during inspiral of the ordinary and superfluid f-modes for the models with and without entrainment (see legend).}
\end{center}
\end{figure}

Evolving the fundamental modes of our three stellar models, we arrive at the results in Figs.~\ref{fig:En1}--\ref{fig:En3}. 
As expected, the ordinary f-mode  is the mode with the largest energy and  dominates the evolution. Moreover, 
the behaviour of the $^2$f$_{\rm o}$-mode does not depend on the entrainment or, indeed, superfluidity. 
As is evident from Fig.~\ref{fig:En3}, the mode evolution is  barely distinguishable in the two models. For the  `normal' star we find practically the same result (not shown in the figure).  

Figure~\ref{fig:En1} shows that the superfluid f-mode is  excited during the inspiral and we note that, in this case, entrainment has a relevant impact. In general, the energy of this mode is more than two orders of magnitude larger in the  model with entrainment. Again, 
this is due to the large effective mass which couples more efficiently the two degrees of freedom and leads to an enhancement of the overlap integral. However, even in the most promising case the 
mode energy of the superfluid f-mode is about two orders of magnitude smaller than that of the ordinary f-mode. Still, it is worth noting that, in the model with strong entrainment
the $^2$f$_{\rm s}$-mode dominates the contribution from
the  pressure, crustal and interface modes.
\subsection{The crust breaking}

Finally, let us consider  the stress built up in the crust by the resonant oscillation modes. 
 Crust failure depends on the detailed physics of the crust and represents, in general, a very complex problem. However, 
we can estimate the breaking strain with the von Mises criterion which is based 
on the elastic strain, $\bar \sigma_{ i j} = \sigma_{ i j} / \check \mu $, 
where the stress tensor is given in Eq.~(\ref{eq:sigij}). 
A key difference with `normal' matter stars is that the Lagrangian displacement in the inner crust 
is given only by the neutral conglomerate of 
charged particles. However, moving towards the neutron drip transition, the mode eigenfunction $\xi_{\ch}^i$ tends to the 
co-moving $\xi^i$, as expected. 
Crust fracture is established by comparing 
\begin{equation}
    \bar \sigma = \sqrt{\frac{1}{2} \bar \sigma_{ i j}^{\ast} \bar \sigma^{ i j} } \label{eq:sigbar}
\end{equation}
to the assumed strain breaking $\bar \sigma_b$. Different models in the literature suggest 
$\bar \sigma _b = 0.1$ \citep{HK09} and $\bar \sigma _b = 0.04$ \citep{BC18} as the breaking limit \citep[see][for more details]{PAP21}. 

\begin{figure*}
\begin{center}
\includegraphics[height=50mm]{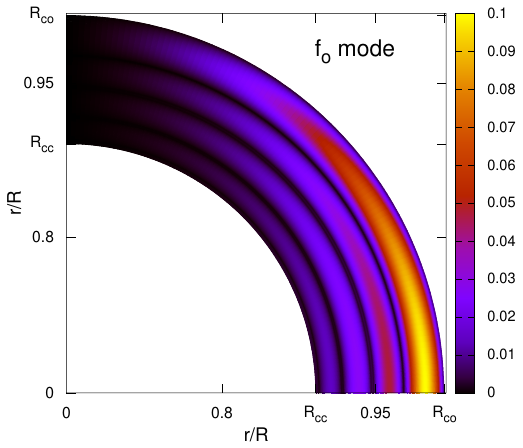}
\includegraphics[height=50mm]{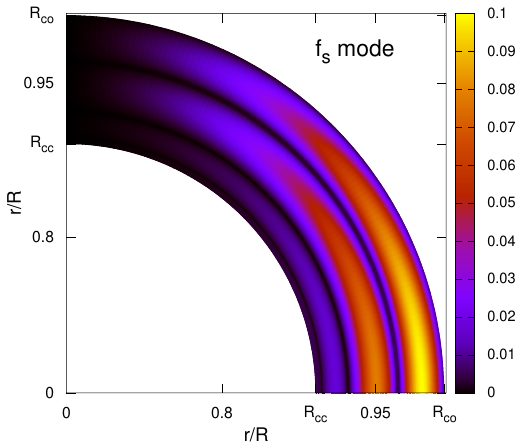}
\includegraphics[height=50mm]{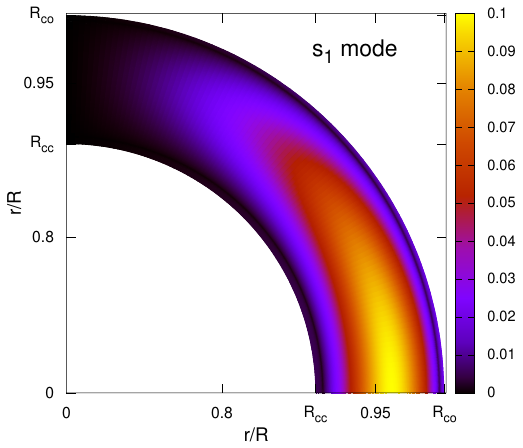}
\caption{ \label{fig:break}
Strain field $\bar \sigma $ of three  modes 
which reach the breaking limit, $\bar \sigma_b =0.1$, during inspiral. 
For a superfluid model with entrainment we show the 
meridional 2D cross-sections of the ordinary f-mode (left panel), the superfluid f-mode (middle panel) and the first crustal shear mode, s$_1$-mode (right panel).  
Lighter colours indicate larger strain.}
\end{center}
\end{figure*}
For each oscillation mode we determine the energy required to reach 
$\bar \sigma = \bar \sigma_b$ at some point (or rather, some region) of the crust. 
We denote this energy by $E_b$ and report the results 
for the case $\bar \sigma_b=0.1$ in Tables \ref{tab:Eb-sf}-\ref{tab:Eb1D} for our three models.  For the case $\bar \sigma_b=0.04$, one can easily determine the breaking energy from a simple rescaling $E_b |_{\bar \sigma = 0.04} = 0.16 E_b |_{\bar \sigma = 0.1}$ \citep{PAP21}.
 
Not all  oscillation modes reach the energy required to fracture the crust during the binary evolution. The outcome depends entirely on the detailed properties of the  modes. 
In Figs.~\ref{fig:En1} and \ref{fig:En2} we indicate, for our stellar models,  the breaking point for the most relevant  modes. 
The interface mode associated with the crust-ocean transition, i$_1$, is the first to reach the breaking limit, at about $D\simeq 360$km. This is a low frequency mode which has time to build up stresses on the crust during the early phases of inspiral, when the binary evolution is slow. In this case the crust cracking is strongly localised near the crust-ocean transition where the eigenfunctions have a characteristic cusp. 
The next resonant mode is the i$_2$-mode, arising due to the core-crust transition, but it does not reach the breaking limit in our models. 
Although the fundamental modes do not reach resonance with the orbital motion, they are significantly amplified close to merger, to the point where they may fracture the crust. 
As previously mentioned, the ordinary f-mode dominates the other modes and reaches the breaking limit at about $D \simeq 80-120$km---depending on the model and $\bar \sigma_b$ (see Figs.~\ref{fig:En1} and \ref{fig:En2}). 
From these figures we also note that the superfluid f-mode can fracture the crust only for  strong entrainment. The crust then fractures  
when $D \simeq 70-100$km. 
In the late inspiral phase, just before  merger,  the first and second shear modes may also  break the crust.

\begin{table}
\begin{center}
\caption{\label{tab:Eb-sf} Mode excitation and breaking energy for the superfluid stellar model without entrainment. We provide the maximum resonant energy 
$E_\mathrm{max}$ (second column) and  the breaking energy $E_b$  (third column; normalised to $E_0 = G M_\star^2 /R$) for the most relevant modes (first column). 
The breaking energy is determined from the von Mises criterion for a breaking strain $\bar \sigma_b = 0.1$}
\begin{tabular}{c  c c   }
\hline
  Mode & $   E_{\rm max} / E_0 $  &  $ E_{b} / E_0 $  \\ 
            &                                        &  $ \bar \sigma_{b} =0.1 $  \\  
\hline
f$_{\rm o}$ & 	$1.59\times 10^{-1}$ &  $ 2.79\times 10^{-6} $	   \\ 
f$_\s$	&  $1.87\times 10^{-6}$  &      $ 3.48\times 10^{-6  } $\\ 
i$_2$ 	& 	\hspace{0.5mm} $1.21\times 10^{-11}$ &	$3.36\times 10^{- 9} $ \\ 
i$_1$ 	& 	\hspace{1.1mm}$9.59\times 10^{-13}$ &	\hspace{0.5mm}	$2.07\times 10^{- 13} $ \\
s$_1$	& 	$4.45\times 10^{-9}$  &	$ 1.94\times 10^{-9}$ \\ 
s$_2$	& 	$2.93\times 10^{-9}$  &	\hspace{0.7mm}$ 7.94\times 10^{-10} $ \\
\hline   
\end{tabular}
\end{center}
\end{table}

\begin{table}
\begin{center}
\caption{\label{tab:Eb-Ch} Same as Table~\ref{tab:Eb-Ch}, but for the superfluid model with entrainment. }
\begin{tabular}{c  c c   }
\hline
  Mode & $   E_{\rm max} / E_0 $  &  $ E_{b} / E_0 $  \\ 
            &                                        &  $ \bar \sigma_{b} =0.1 $  \\  
\hline
f$_{\rm o}$ & 	$1.74\times 10^{-1} $  &      $ 1.46\times 10^{-6} $	   \\ 
f$_\s$	& 	$8.47\times 10^{-4} $  &      $ 3.44\times 10^{-8} $\\ 
i$_2$ 	& \hspace{0.5mm}	$2.88\times 10^{- 12} $  & 	$1.52\times 10^{- 9} $ \\ 
i$_1$ 	& \hspace{0.5mm}	$1.02\times 10^{- 12} $ &	\hspace{0.5mm}	$2.07\times 10^{- 13} $ \\
s$_1$	& 	$2.61\times 10^{-8}$  &	$ 2.95\times 10^{-9}$ \\ 
s$_2$	& 	$1.55\times 10^{-9}$  &	$ 1.25\times 10^{-9} $ \\
\hline   
\end{tabular}
\end{center}
\end{table}

\begin{table}
\begin{center}
\caption{\label{tab:Eb1D} Same as Table~\ref{tab:Eb1D}, but for the single-fluid model.}
\begin{tabular}{c  c c   }
\hline
  Mode & $   E_{\rm max} / E_0 $  &  $ E_{b} / E_0 $  \\ 
            &                                        &  $ \bar \sigma_{b} =0.1 $  \\  
\hline
f	        & 	\hspace{-0.4mm}$1.56\times 10^{-  1}$      &      $ 6.41\times 10^{-6} $	   \\ 
i$_2$ 	& \hspace{0.4mm}	$4.02\times 10^{-  12} $ &	\hspace{0.5mm}	$3.79\times 10^{- 10} $ \\ 
i$_1$ 	& \hspace{0.5mm}	$6.52\times 10^{-  13} $ &	\hspace{0.5mm}	$2.11\times 10^{- 13} $ \\
s$_1$	& 	$1.73\times 10^{-9}$  &	$ 3.24\times 10^{-9}$ \\ 
s$_2$	& 	$2.26\times 10^{-9}$  &	$ 1.38\times 10^{-9} $ \\
\hline   
\end{tabular}
\end{center}
\end{table}


\section{Concluding remarks}

When they reach the late stages of binary inspiral, neutron stars should be sufficiently mature to host superfluid 
and superconducting constituents throughout the fluid core and free superfluid neutrons in the inner crust. Hence, it is natural to consider the impact, if any, of superfluid physics on the tidal 
response of a neutron star. In order to explore this issue, we  extended our previous work \citep{PAP21}, 
using the two-fluid formalism to account for the 
superfluid degrees of freedom. The star was taken to be a Newtonian spherically symmetric body with a core, 
inner and outer crust and an ocean.  We considered a barotropic two-fluid equation of state and 
 models with composition gradients in the core and 
crust which mimic the expected values for  realistic equations of state.  In our models we 
also introduced entrainment representing the large effective neutron mass expected at the bottom of the inner crust. 
With this set up we confirmed that entrainment may have significant impact on the mode frequencies of the superfluid fundamental and pressure modes \citep{pass}. 

We explored the tidal response in both the static limit and the dynamical case, evolving the mode amplitude excited by the tidal driving. 
In the static limit, we showed that the difference in the Love number between models with and without superfluid 
constituents is not significant (a result that accords with a general argument that superfluidity should not affect the static tide). As for normal stars, the main contribution to the 
static deformation is given by the fundamental mode. In a superfluid model with strong entrainment, we noted
 a slight decrease of the effective Love number contribution from the ordinary f-mode, compensated for by an increased contribution from the superfluid f-mode. The latter is still likely too small to be distinguishable by observations, but the result demonstrates that we need to consider the complete set of oscillation modes in order to establish the precise tidal response. This agrees with the demonstrations from \citet{ap20a} and \citet{PAP21}.

The overall result for the tidal excitation is very similar to stars with `normal' matter. Basically, even though we show  that the superfluid f-mode energy may be larger than that of the shear and interface modes, it is always much smaller than the ordinary f-mode.  In the most favourable case, the contribution of the superfluid f-mode is at the percent level.

We also consider the mode-induced breaking of the crust. Our results are similar to those of \citet{PAP21}, although it is worth noting that the superfluid f-mode may also induce crust fracture before the eventual merger of a neutron star binary. This is, at least conceptually, interesting. It is also worth noting that different regions in the crust are strained by different modes, see Fig.~\ref{fig:break}, so even if the normal fundamental mode has exceeded the breaking strain, there may still be scope for  higher density crust failures to be induced by the superfluid f-mode. This is, of course, speculative given that we do not really know what happens once the crust reaches the breaking strain in an extended region. If it is a local failure, then our suggestion might have some merit. If it is a global event then this seems less likely.

Qualitatively, our results should be expected to carry over to more realistic, relativistic models based on nuclear equations of state with detailed superfluid pairing gaps and entrainment. Quantitatively, the results may (obviously) change. We need to be mindful of the fact that the use of a realistic equation of state (within a fully relativistic calculation, following, for example, \citealt{grmodes}) may shift the results. We also need to pay attention to the fact that aspects of the entrainment are still being discussed \citep{ent1,ent2,ent3,ent4,ent5}. At the end of the day, the superfluid imprint on the tide is likely to remain small. Our results suggest that a detection of the effect would require percent level accuracy in the parameter inference, which will be a challenge even for third generation gravitational-wave instruments like the Einstein Telescope and the Cosmic Explorer. Of course, the conclusions need to be confirmed with fully  relativistic models. The first step in this direction would be to develop the mode-sum strategy in relativity. So far there has not been much (if any) progress in this respect. Work in this direction should clearly be encouraged.


\section*{Acknowledgments}

NA  is grateful for support from STFC via grant numbers ST/R00045X/1 and ST/V000551/1. PP acknowledges support from the `Ministero dell'istruzione, dell’universit\`a e della ricerca' (MIUR) PRIN 2017 programme (CUP: B88D19001440001) and from the Amaldi Research Center funded by the MIUR programme `Dipartimento di Eccellenza' (CUP: B81I18001170001). 


\subsection*{Data Availability}

The main data underlying this article are available in the article. Additional information may be requested from the authors.



\begin{thebibliography}{}

\bibitem[\protect\citeauthoryear{Allard and Chamel}{2021}]{ent5}
Allard, V., Chamel, N., 2021, Universe, 7, 470

\bibitem[\protect\citeauthoryear{Andersson}{2021}]{NA21} Andersson, N., 2021, 
Universe,  7, 17
 
\bibitem[\protect\citeauthoryear{Andersson and Comer}{2001}]{AC01}
Andersson N., Comer G.L., 2001, MNRAS, 328, 1129

\bibitem[\protect\citeauthoryear{Andersson, Comer and Grosart}{2004}]{kirsty}
Andersson N., Comer G.L., Grosart, K., 2004, MNRAS,  355,  918

\bibitem[\protect\citeauthoryear{Andersson, Glampedakis and Haskell}{2009}]{fmode2}
Andersson, N., Glampedakis, K, Haskell, B, 2009, 
Phys. Rev. D,  79, 103009

\bibitem[\protect\citeauthoryear{Andersson et al}{2012}]{crust1}
Andersson, N., Glampedakis, K., Ho, W.C.G, Espinoza, C.M., 2012, Phys. Rev. Lett.,  109, 241103


\bibitem[\protect\citeauthoryear{Andersson, Haskell and Samuelsson}{2011}]{lagrange}
Andersson, N., Haskell, B. Samuelsson, L., 2011, MNRAS, 416, 118

\bibitem[\protect\citeauthoryear{Andersson and Pnigouras}{2019}]{ap19}
Andersson  N., Pnigouras P., 2019,  MNRAS, 489, 4043

\bibitem[\protect\citeauthoryear{Andersson and Pnigouras}{2020}]{ap20a}
Andersson  N., Pnigouras P., 2020,  Phys. Rev. D,  101, 083001

\bibitem[\protect\citeauthoryear{Andersson and Comer}{2021}]{LivRev}
Andersson N., Comer G.L., 2021, 
Living Reviews in Relativity, 24, 3

\bibitem[\protect\citeauthoryear{Ashton et al}{2019}]{ash}
Ashton, G., Lasky, P.D., Graber, V., Palfreyman, J., 2019, Nature Astronomy,  3, 1143

\bibitem[\protect\citeauthoryear{Baiko and Chugunov}{2018}]{BC18}
Baiko, D.A., and A.I. Chugunov, 2018, MNRAS, 480, 5511

\bibitem[\protect\citeauthoryear{Carter, Chamel and Haensel}{2005}]{Carter05}
Carter B., Chamel N., Haensel P., 2005, Nucl. Phys. A, 748, 675

\bibitem[\protect\citeauthoryear{Chamel}{2005}]{Ch05}
Chamel N., 2005, Nucl. Phys. A, 747, 109

\bibitem[\protect\citeauthoryear{Chamel}{2006}]{Ch06}
Chamel N., 2006, Nucl. Phys. A, 773, 263

\bibitem[\protect\citeauthoryear{Chamel}{2013}]{crust2}
Chamel, N., 2013, Phys. Rev. Lett., 110, 011101

\bibitem[\protect\citeauthoryear{Comer, Langlois and Lin}{1999}]{comer}
Comer, G.L. Langlois, D., Lin. L.-M., 1999, Phys. Rev. D, 60, 104025

\bibitem[\protect\citeauthoryear{Datta and Char}{2020}]{Datta}
Datta S., Char P., 2020, Phys. Rev. D 101, 064016

\bibitem[\protect\citeauthoryear{Delsate et al}{2016}]{ent2}
Delsate, T., Chamel, N.; Gurlebeck, N., Fantina, A.F., Pearson, J.M., Ducoin, C., 2016,  Phys. Rev. D  94, 023008

\bibitem[\protect\citeauthoryear{Douchin and Haensel}{2001}]{douchin}
Douchin F., Haensel P., 2001,  Astron. Astrop., 380, 151

\bibitem[\protect\citeauthoryear{Epstein}{1988}]{epstein}
Epstein, R.~I., 1988, Ap. J., 333, 880

\bibitem[\protect\citeauthoryear{Friedman and Schutz}{1978}]{fs78}
Friedman, J.L., Schutz, B.F., 1978, Ap. J., 221, 937

\bibitem[\protect\citeauthoryear{Graber, Cumming and Andersson}{2018}]{grab}
Graber, V., Cumming, A., Andersson, N., 2018, Ap. J.,  865, 23

\bibitem[\protect\citeauthoryear{Gualtieri et al.}{2014}]{Gua14}
Gualtieri L., Kantor E. M. , Gusakov M. E., Chugunov A. I., 2014,  Phys. Rev. D, 90, 2

\bibitem[\protect\citeauthoryear{Gusakov and Kantor}{2011}]{GK11}
Gusakov M. E., Kantor E. M., 2011, Phys. Rev. D, 83, 8

\bibitem[\protect\citeauthoryear{Gusakov and Kantor}{2013}]{GK13}
Gusakov M.~E., Kantor E.~M., 2013, Phys. Rev. D, 88, 101302

\bibitem[\protect\citeauthoryear{Haskell and Melatos}{2015}]{bryn}
Haskell, B., Melatos, A., 2015, Int. J. Mod. Phys. D, 24, 1530008

\bibitem[\protect\citeauthoryear{Horowitz and Kadau}{2009}]{HK09}
Horowitz, C.J., and K. Kadau, 2009, Phys. Rev. Lett. 102, 191102

\bibitem[\protect\citeauthoryear{Kantor and Gusakov}{2011}]{KG11}
Kantor E. M., Gusakov M. E., 2011, Phys. Rev. D, 83, 10

\bibitem[\protect\citeauthoryear{Kantor and Gusakov}{2014}]{KG14}
Kantor E.~M., Gusakov M.~E., 2014, MNRAS, 442, L90

\bibitem[\protect\citeauthoryear{Lai}{1994}]{Lai}
Lai D., 1994, MNRAS, 270, 611

\bibitem[\protect\citeauthoryear{Lee}{1995}]{Lee}
Lee U., 1995, Astron. Astrophys., 303, 515

\bibitem[\protect\citeauthoryear{Lin, Andersson and Comer}{2008}]{grmodes}
Lin, L.-M., Andersson, N., Comer, G.L., 2008, Phys. Rev. D,  78,  083008

\bibitem[\protect\citeauthoryear{Lindblom and Mendell}{1995}]{fmode1}
Lindblom, L., Mendell, G., 1995, Ap. J., 444, 804

\bibitem[\protect\citeauthoryear{Mendell}{1991a}]{mend1}
Mendell, G., 1991a, Ap. J., 380, 515

\bibitem[\protect\citeauthoryear{Mendell}{1991b}]{mend2}
Mendell, G., 1991b, Ap. J., 380, 530

\bibitem[\protect\citeauthoryear{Noel and Urban}{2016}]{ent1}
Noel, M., Urban, M., 2016, Phys. Rev. C, 94, 065801

\bibitem[\protect\citeauthoryear{Page et al}{2011}]{page}
Page, D., Prakash, M., Lattimer, J.M, Steiner, A.W., 2011,
Phys. Rev. Lett. 106, 081101

\bibitem[\protect\citeauthoryear{Palfreyman et al}{2018}]{palf}
Palfreyman, J., Dickey, J.M., Hotan, A., Ellingsen, S., van Straten, W., 2018, Nature,  556, 219

\bibitem[\protect\citeauthoryear{Passamonti and Andersson}{2012}]{pass}
Passamonti A., Andersson N., 2012, MNRAS 419, 638

\bibitem[\protect\citeauthoryear{Passamonti, Andersson and Pnigouras}{2021}]{PAP21}
Passamonti A., Andersson N.,  Pnigouras P., 2021,  MNRAS 504, 1273

\bibitem[\protect\citeauthoryear{Passamonti, Andersson and Ho}{2016}]{PAH16}
Passamonti A., Andersson N.,  Ho W.~C.~G., 2016,  MNRAS 455, 1489

\bibitem[\protect\citeauthoryear{Poisson and Will}{2014}]{PWbook}
Poisson, E., Will, C.M., 2014, {\em Gravity}, (Cambridge University Press, Cambridge)

\bibitem[\protect\citeauthoryear{Prix, Comer and Andersson}{2002}]{Prix}
Prix R., Comer G. L., Andersson N., 2002, Astron. Astrophys., 381, 178

\bibitem[\protect\citeauthoryear{Prix}{2004}]{Prix04}
Prix R., 2004, Phys. Rev. D, 69, 043001

\bibitem[\protect\citeauthoryear{Prix and Rieutord}{2004}]{PR02}
Prix, R., Rieutord, M., 2002, Astron. Astrophys., 393, 949

\bibitem[\protect\citeauthoryear{Rau and Wasserman}{2018}]{RW18}
Rau, P. B., Wasserman I., 2018, MNRAS 481, 4427

\bibitem[\protect\citeauthoryear{Sauls, Chamel and Alpar}{2020}]{ent4} 
Sauls, J.A., Chamel, N., Alpar, M.A., 2020, {\em Superfluidity in Disordered Neutron Stars Crusts}, preprint arXiv:2001.09959.

\bibitem[\protect\citeauthoryear{Shternin et al}{2011}]{shtern1}
Shternin, P.S., Yakovlev, D.G., Heinke, C.O., Ho, W.C.G., Patnaude, D.J., 2011, MNRAS Letters,  412,  L108

\bibitem[\protect\citeauthoryear{Shternin et al}{2021}]{shtern2}
Shternin, P.S., Ofengeim, D.D., Ho, W.C.G., Heinke, C.O., Wijngaarden, M.J.P., Patnaude, D.J., 2021, MNRAS,  506, 709

\bibitem[\protect\citeauthoryear{Watanabe and Pethick}{2017}]{ent3}
Watanabe, G., Pethick, C.J., 2017, 
Phys. Rev. Lett., 119, 062701

\bibitem[\protect\citeauthoryear{Yeung et~al}{2021}]{Yeung}
Yeung C.-H., Lin L.-M., Andersson N., Comer G. L., 2021, Universe, 7,  111

\bibitem[\protect\citeauthoryear{Yoshida and Lee}{2003}]{YL03}
Yoshida S., Lee U., 2003, MNRAS 344, 207

\bibitem[\protect\citeauthoryear{Yu and Weinberg}{2017}]{wein}
Yu H., Weinberg N., 2017, MNRAS, 464, 2622

\end{thebibliography}
\end{document}